\documentclass[11pt, a4paper]{article}

\usepackage[margin=1in]{geometry} % Standard margins
\usepackage{setspace}
\usepackage{amsmath, amssymb, amsfonts}
\usepackage{bm}        % Bold math
\usepackage[version=4]{mhchem}    % Chemical formulas
\usepackage{siunitx}   % Physics units
\usepackage{graphicx}
\usepackage{float}
\usepackage{booktabs}

\usepackage[T1]{fontenc}
\usepackage[utf8]{inputenc}
\usepackage{authblk}

\usepackage[colorlinks=true, linkcolor=blue, citecolor=blue, urlcolor=blue]{hyperref}

\usepackage[numbers,sort&compress]{natbib}

\title{Development and Application of an eV Neutron Polarization for Parity Violation Studies at CSNS Back-n Beamline}

\author[1,3]{Xu Qin}
\author[2,3,5]{Tianhao Wang}
\author[6]{Xuanbo Chen}
\author[2,3]{Changdong Deng}
\author[7]{Yongce Gong}
\author[3,8]{Zenghang Huang}
\author[2,3]{Wei Jiang}
\author[3,9]{Zhengquan Liu}
\author[6]{Guangyuan Luan}
\author[6]{Haotian Luo}
\author[2,3,4]{Qiuyue Luo}
\author[2,3]{Yongjia Lv}
\author[2,3]{You Lv}
\author[2,3]{Nikolaos Vassilopoulos}
\author[6]{Xichao Ruan}
\author[10]{William Michael Snow}
\author[2,3]{Kang Sun}
\author[10]{Sepehr Samiei}  
\author[2,3,4]{Jian Tang}
\author[3,11]{Shilin Wang}
\author[6]{Hongyi Wu}
\author[1]{Xiaomin Xiong}
\author[2,3,12]{Xinyu Yuan}
\author[2,3,5]{Junpei Zhang}
\author[3,10]{Mofan Zhang}
\author[6]{Qiwei Zhang}
\author[2,3,4]{Qingbo Zheng}

\author[2,3]{Ruirui Fan\thanks{Corresponding author: fanrr@ihep.ac.cn}}
\author[2,3,4]{Xin Tong\thanks{Corresponding author: tongxin@ihep.ac.cn}}

\affil[1]{Center for Neutron Science and Technology, Guangdong Provincial Key Laboratory of Magnetoelectric Physics and Devices, School of Physics, Sun Yat-Sen University, Guangzhou, Guangdong 510275, China}
\affil[2]{Institute of High Energy Physics, Chinese Academy of Sciences, Beijing 100049, China}
\affil[3]{Spallation Neutron Source Science Center, Dongguan 523803, China}
\affil[4]{University of Chinese Academy of Sciences, Beijing, 100049, China}
\affil[5]{Guangdong Provincial Key Laboratory of Extreme Conditions, Dongguan 523803, China}
\affil[6]{National Key Laboratory of Nuclear Data, China Institute of Atomic Energy, Beijing 102413, China}
\affil[7]{School of Physical Sciences, Great Bay University, Dongguan, Guangdong 523000, China}
\affil[8]{National Synchrotron Radiation Laboratory, University of Science and Technology of China, Hefei, Anhui 230026, China}
\affil[9]{School of Mechanical Engineering of DGUT, Dongguan University of Technology, Dongguan, Guangdong 523808, China}
\affil[10]{Indiana University, Bloomington, IN 47405, USA}
\affil[11]{School of Physics, MOE Key Laboratory for Nonequilibrium Synthesis and Modulation of Condensed Matter, Xi'an Jiaotong University, Xi'an, Shaanxi 710049, China}
\affil[12]{Shandong Provincial Key Laboratory of Nuclear Science, Nuclear Energy Technology and Comprehensive Utilization, Weihai Frontier Innovation Institute of Nuclear Technology, School of Nuclear Science, Energy and Power Engineering, Shandong University, Shandong 250061, China}

\date{} % No date for arXiv

\begin{document}

\maketitle

\begin{abstract}
The dynamic enhancement of symmetry-breaking effects in neutron-nucleus resonances provides a sensitive testing ground for Time-Reversal Invariance Violation (TRIV). Exploiting this mechanism, the Neutron Optics Parity and Time Reversal Experiment (NOPTREX) seeks to elucidate the origin of the universe's baryon asymmetry. Critical to this effort is the precise measurement of Parity Violation (PV) asymmetries, which is essential to calibrate the nuclear parameters required for future TRIV experiments. To facilitate these studies, we developed an eV polarized neutron at the Back-n white neutron beamline of the China Spallation Neutron Source (CSNS). Neutron polarization is generated by an in-situ Spin-Exchange Optical Pumping (SEOP) \ce{^3He} filter. Spin manipulation is performed by an adiabatic spin flipper, while spin polarization is preserved over the flight path by a vacuum transport system equipped with a solenoidal guide field. Experiments successfully measured an asymmetry of approximately $7.8 \pm 2.4~(\text{stat.}) \pm 0.3~(\text{sys.})\%$ at the \SI{0.747}{eV} p-wave resonance of \ce{^{139}La}. These results are in agreement with previous results on this resonance and validate the system’s capability for PV measurements.
\end{abstract}

\vspace{1em}

% --- Main Text Starts Here ---

\section{Introduction}\label{section1}

The origin of the Baryon Asymmetry of the Universe (BAU) remains a central question in cosmology~\cite{shaposhnikov2009baryon,kobayashi1973cp}. According to the Sakharov criteria, generating the observed matter-antimatter asymmetry requires baryon number violation, departure from thermal equilibrium, and C- and CP-violation~\cite{sakharov1998violation, canetti2012matter}. However, the CP violation incorporated in the Standard Model (SM) via the CKM matrix is insufficient to explain the observed baryon-to-photon ratio, with the discrepancy estimated to be as large as ten orders of magnitude. This discrepancy motivates the search for Beyond the Standard Model (BSM) sources of CP violation~\cite{gavela1994standard}. Under the assumption of CPT conservation, Time-Reversal Invariance Violation (TRIV) is equivalent to CP violation. Consequently, precision searches for TRIV in nuclear systems have become a priority. However, direct detection of TRIV is experimentally challenging due to the minute magnitude of the effect. To overcome this, we utilize the specific enhancement mechanisms found in neutron-nucleus interactions.

This pursuit relies on the critical insight that the nuclear matrix elements enhancing Parity Violation (PV) similarly amplify time-reversal effects; thus, the sensitivity to TRIV scales directly with the magnitude of the PV asymmetry. Compound nuclear resonances offer a powerful physical amplification mechanism at low neutron energies for addressing this challenge. Seminal work by Sushkov and Flambaum established that the mixing of $s$-wave (orbital angular momentum $l=0$) and $p$-wave ($l=1$) resonances through weak interaction enhances parity-violating amplitudes by factors of $10^4$ to $10^6$ in the vicinity of $p$-wave resonances~\cite{sushkov1980possibility}.

We can write the total angular momentum $j$ of incident neutrons as
\begin{equation}
    j = l + s
\end{equation}
where $l$ is the orbital angular momentum of the incident neutron with respect to the nucleus center of mass, and $s=1/2$ is the neutron spin. The longitudinal asymmetry, often reported in the literature~\cite{hayes2001parity} as a measure of the size of the PV effect, has a theoretical approximate form:
\begin{align}
A_L = 2 \sum_{s} \frac{W}{E_s - E_p} \sqrt{\frac{\Gamma_s^n}{\Gamma_p^n}}
\end{align}
where $E_p$ and $E_s$ are the $p$- and $s$-wave resonance energies, $W = \langle p | W_{PV} | s \rangle$ is the weak matrix element transitioning the neutron wave function from the $s$-wave ($p$-wave) state to the $p$-wave ($s$-wave) state. $\Gamma^n_s = \Gamma^n_{s, j=1/2}$ and $\Gamma^n_p = \Gamma^n_{p, j=1/2} + \Gamma^n_{p, j=3/2}$ are the $s$- and $p$-wave resonance total neutron widths, and $x$ is the ratio of the $j=1/2$ $p$-wave partial neutron width to the total neutron width, defined as $x = \sqrt{\Gamma^n_{p, j=1/2} / \Gamma^n_p}$. The $p$-wave amplitude can be in the $j=1/2$ or $j=3/2$ state, whereas the $s$-wave amplitude only allows the $j=1/2$ state. The $s$- and $p$-wave resonances of opposite parities can mix through weak interaction when the total angular momentum $J$ of the two states is equal.

\begin{figure}[t] 
    \centering
    \includegraphics[width=\linewidth]{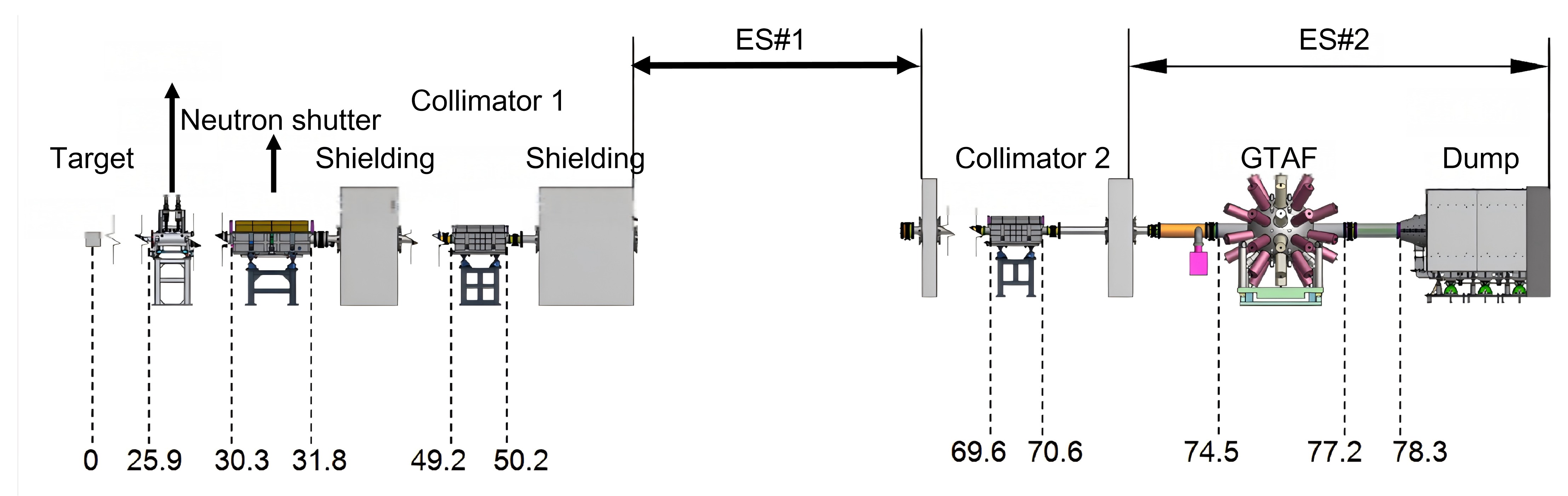} 
    \caption{Layout of the Back-n beam line at CSNS.}
    \label{fig:Back-n}
\end{figure}

Having established the mechanism for PV enhancement, it is necessary to formally connect this P-odd effect to the T-odd observable. This giant amplification arises from the interplay of two distinct factors: a dynamical enhancement resulting from sufficiently large weak matrix elements coupled with the small energy spacing ($E_s - E_p$) between the high-density compound states, and a kinematical enhancement resulting from the admixture of a large $s$-wave neutron width amplitude ($\sqrt{\Gamma_n^s}$) into a naturally small $p$-wave amplitude ($\sqrt{\Gamma_n^p}$). Each mechanism produces an estimated enhancement of $10^2$ to $10^3$.
For $I=1/2$ nuclei, the forward scattering amplitude $f$ can be written as the following for terms up to vector polarization:
\begin{equation}
    f = f_0 + f_1 \bm{\sigma} \cdot \bm{I} + f_2 \bm{\sigma} \cdot \bm{k} + f_3 \bm{\sigma} \cdot (\bm{I} \times \bm{k})
\end{equation}
where $\bm{\sigma}$, $\bm{I}$, and $\bm{k}$ are respectively the spin of the incident neutron, the spin of the target nuclei, and the momentum of the neutron. In the above expression, $f_0$ corresponds to the neutron absorption cross section, and $f_1$ is the pseudo-magnetic term ~\cite{ bowman2014search}. Both $f_0$ and $f_1$ are parity-even and time-reversal-even. $f_2$ is a parity-violating, time-reversal-conserving term, enhanced through the $sp$-mixing mechanism discussed before. $f_3$ is the parity- and time-reversal-violating term, which was predicted to also be enhanced through $sp$-mixing~\cite{gudkov1992cp}. A non-zero $f_3$ then implies T-violation. The P-odd T-odd cross section associated with $f_3$ can be related to the P-odd T-even cross section associated with $f_2$ by the following expression:
\begin{equation}
    \Delta \sigma_{PT} = \kappa(J) \frac{w}{v} \Delta \sigma_P
\end{equation}
where $\Delta \sigma_{PT}$ and $\Delta \sigma_P$ are the T- and P-violating cross sections, respectively; $\kappa(J)$ is a spin factor; and $w$ and $v$ are the T- and P-violating matrix elements. The above relationship between TRIV and the PV effect in the compound nucleus is produced by the recombination between the channel spin $S = s + I$ and $l$ in the same $p$-wave resonance. A detailed derivation can be found in Ref.~\cite{gudkov2018nuclear}. Through the measurement of $\Delta \sigma_P$ and $\kappa(J)$, we can estimate the size of possible T-violation in a $p$-wave resonance of target nuclei. However, $\Delta \sigma_P$ and $\kappa(J)$ can only be determined experimentally due to the extremely complex wavefunction of the compound nucleus.

The Neutron Optics Parity and Time Reversal Experiment (NOPTREX) was formed to address this specific experimental ~\cite{adhikary2025review}. NOPTREX targets these enhanced regimes to conduct a sensitive search for T-violation. Consequently, a comprehensive spectroscopic campaign of PV asymmetries is a prerequisite; these measurements quantify the nuclear amplification factors---specifically the $s$-$p$ mixing parameters---thereby allowing for the identification of optimal target nuclei where the sensitivity to T-violation is maximized. Coordinating this effort, the NOPTREX collaboration builds upon the foundational surveys performed by the TRIPLE collaboration at LANSCE, as well as studies at KEK and Dubna, to systematically screen resonances before proceeding to the high-precision TRIV search. This investigation centers on analyzing the neutron-nucleus forward scattering amplitude, which contains the signature of symmetry breaking.

The TRIPLE Collaboration validated this enhancement mechanism in the 1990s~\cite{bowman1990parity, yuan1991parity, zhu1992parity, frankle1992parity, bowman1993recent, mitchell2001parity}. Through neutron transmission, these measurements revealed parity-violation effects in 75 resonances across 18 nuclides, and further calibrated neutron transmission asymmetry over a larger energy range (0.7--1000~eV) through optimization and upgrades of the equipment. Statistical analysis of these results demonstrated that the observed effects align with theoretical predictions for nucleon-nucleon weak interactions.

Emerging in the 1990s, a distinct experimental technique was developed to significantly improve the accuracy of parity violation measurements. In addition to neutron transmission, methods utilizing $\gamma$-rays emitted from excited states have also been employed to study parity violation~\cite{shimizu1993longitudinal, seestrom1999apparatus, skoy1996isotopic, okudaira2018angular,endo2023measurements}. The detection of radiative capture $\gamma$-rays offers a distinct advantage over transmission measurements, as it remains largely unaffected by potential scattering backgrounds. This allows for precise measurements even in systems with significant scattering cross-sections. Measurements of longitudinal asymmetries were carried out in the 0.4--70~eV range for nuclei including $^{81}\text{Br}$, $^{93}\text{Nb}$, $^{108}\text{Pd}$, $^{111}\text{Cd}$, $^{124}\text{Sn}$, $^{139}\text{La}$, and Xe isotopes. Furthermore, this observable facilitates the further analysis of scattering angles in the context of parity violation, serving as a foundation for probing time-reversal violation.

The current NOPTREX collaboration at CSNS extends this effort by utilizing the intense pulsed neutron source to enhance statistical precision and broaden the accessible energy range. The Back-n beamline provides a decisive geometric advantage for NOPTREX through its extended flight path of $\sim$76~m (ES\#2), compared to the $\sim$21.5~m baseline at J-PARC ANNRI~\cite{okudaira2018angular, tang2021back, chen2019neutron}. This increased flight distance naturally enhances the Time-of-Flight (TOF) resolution, facilitating the utilization of the high-energy neutron source. Beyond the geometric factor, the Back-n white neutron source offers a distinct advantage in temporal precision. Unlike ANNRI, which utilizes a moderator where the pulse structure is inevitably broadened by multiple scattering events during the moderation process, the Back-n beamline operates as an unmoderated neutron source. The absence of a moderator eliminates the emission tails caused by scattering, yielding a pristine time structure. This sharp timing, coupled with the long flight path, produces sharper resonance profiles. Such spectral clarity is essential for disentangling the 0.74~eV $p$-wave resonance in $^{139}\text{La}$ from the dominant $s$-wave background. Consequently, systematic uncertainties in the resonance fitting are reduced, directly enhancing the fidelity of the PV asymmetry extraction.

\begin{figure}[t] 
    \centering
    \includegraphics[width=0.9\linewidth]{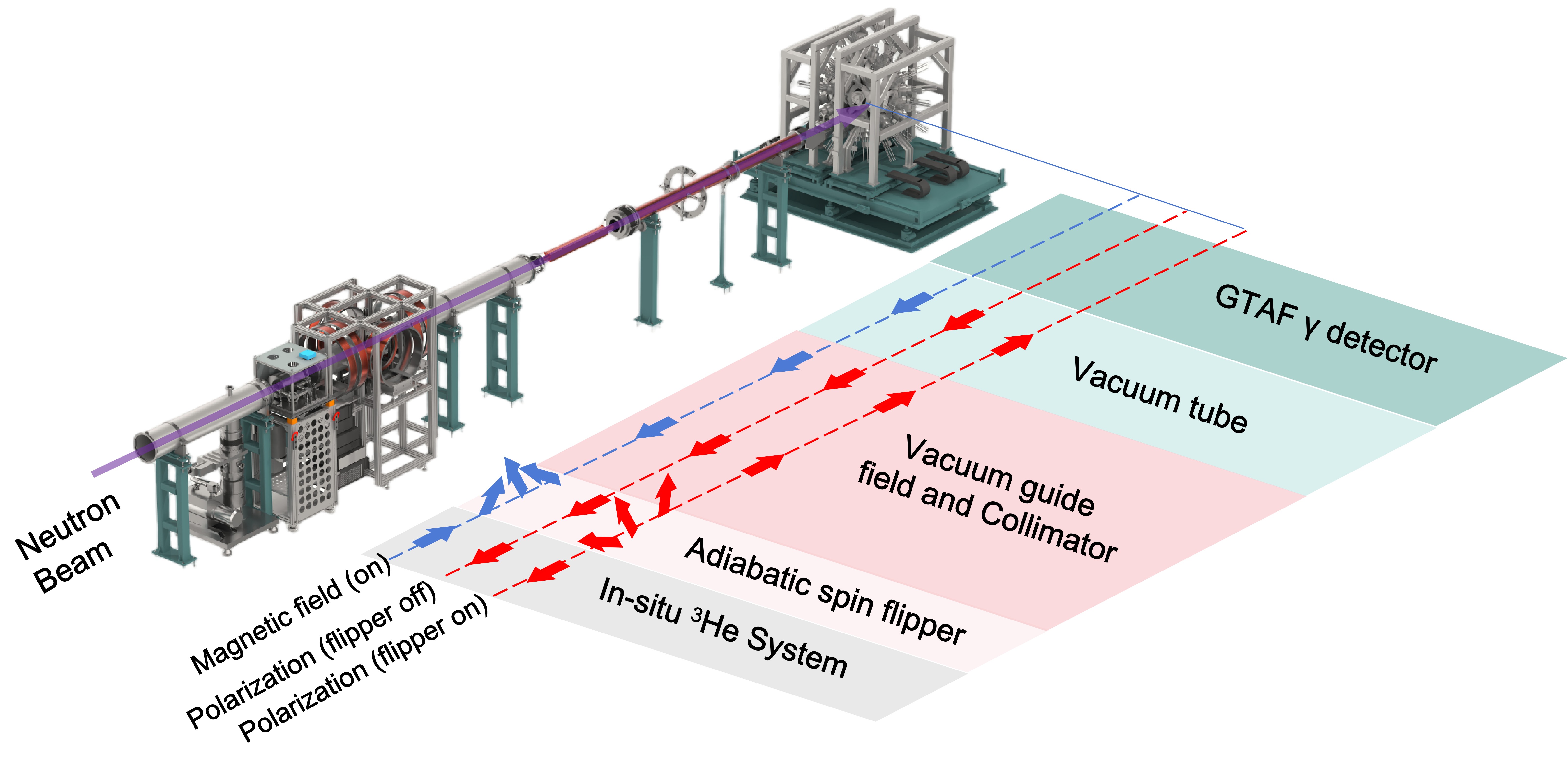}
    \caption{Schematic of the high-energy polarized neutron system, showing the key components such as the \textit{in-situ} \ce{^3He} system, spin flipper, and polarization transport system. These components work in tandem to ensure efficient polarization of neutrons for high-precision parity violation measurements.}
    \label{fig:schematic}
\end{figure}

In this work, we present the experimental apparatus established at the Back-n beamline for precision PV measurements in the eV energy range, and report a measured asymmetry of $7.78\% \pm 2.44\%$ for Lanthanum ($^{139}\text{La}$) at the 0.747~eV resonance. Section~II describes the CSNS Back-n white neutron source, highlighting upgrades implemented specifically for the NOPTREX collaboration. Section~III details the polarized neutron setup, comprising the polarized $^3\text{He}$ target, spin flipper, and magnetic transport guide. Particular emphasis is placed on the custom spin flipper designed to optimize the modulation of eV neutrons, as well as the corresponding electronic circuit design. Experimental results are presented in Section~IV, covering the PV asymmetry measurement, wavelength-dependent spin manipulation efficiency, and an analysis of systematic effects induced by the guide field. Finally, Section~V summarizes the system performance and discusses prospects for future measurements.

\section{Back-n white neutron source}\label{sec:2}

The Back-n white neutron source at the China Spallation Neutron Source (CSNS) is a specialized facility dedicated to high-precision nuclear data measurements~\cite{chen2024measurement}. Delivering an intense, continuous energy spectrum spanning from thermal energies up to \SI{300}{MeV}, Back-n serves as a premier platform for resolving critical neutron-induced reaction cross-sections, facilitating advancements in both nuclear energy technology and fundamental physics~\cite{tang2021back,chen2019neutron}. The facility integrates several key technical advantages: a high neutron flux (peaking at \SI{8.6e6}{n/cm^2/s}), an extended flight path, and low beam divergence. As shown in Fig.~\ref{fig:Back-n}, the flight paths from the spallation target to the experimental stations (\SI{55}{m} for ES\#1 and \SI{77}{m} for ES\#2) provide the Time-of-Flight (TOF) resolution necessary to resolve complex resonance structures. This long neutron flight path, combined with low beam divergence, minimizes background interference and optimizes the signal-to-noise ratio. Such conditions make it possible to meet the stringent precision requirements of parity violation studies, rendering Back-n a suitable venue for the NOPTREX experiment.

To mitigate neutron and gamma backgrounds in ES\#1 and ES\#2, a multi-layered shielding strategy is employed. Key components include a reinforced steel-block barrier separating ES\#1 from the RTBT tunnel, followed by massive concrete shielding downstream of the shutter and collimators. Furthermore, the interior walls and ceiling of ES\#1 are lined with a \SI{50}{mm} layer of borated polyethylene for neutron absorption, and a beam dump is installed at the end of ES\#2.

The collimation system suppresses background and defines the beam size at the Back-n beamline. Two collimators are installed: one upstream of End Station~1 (ES\#1) and a second between ES\#1 and End Station~2 (ES\#2). The NOPTREX experiment used the second collimator, which provides apertures with diameters of \SIlist{20;30;60}{mm}. We selected the \SI{30}{mm} diameter aperture to maintain sufficient neutron flux while limiting the beam size and reducing stray scattering from the downstream polarization transport components. Further downstream at ES\#2, the beamline is equipped with a comprehensive detection system and beam termination apparatus. The station houses $\gamma$-ray detectors and neutron detectors for signal acquisition. Additionally, a beam dump is installed at the end of the flight path to safely terminate the neutron beam and minimize background noise arising from backscattering.

\section{eV neutron polarization}\label{sec:3}
The eV polarized neutron system comprises an in-situ Spin-Exchange Optical Pumping (SEOP) \ce{^3He} system, a neutron spin flipper, and a magnetic guide field. Together with the Gamma Total Absorption Facility (GTAF) and the Back-n beamline, these components constitute the CSNS NOPTREX experimental system. The mechanical layout of the apparatus is illustrated in Fig.~\ref{fig:schematic}. Polarized neutrons are generated via the \ce{^3He} system, with a polarization of approximately \SI{30}{\percent} obtained at \SI{0.74}{eV}. Detector gain may exhibit linear time-dependent drift, primarily driven by factors such as the temperature dependence of electronic components. To suppress this effect, a spin flipper is employed to execute rapid spin reversal, thereby shortening the measurement interval between distinct spin states to \SI{0.4}{s}~\cite{schaper2020modular, matsubayashi2015correction}. To preserve the neutron spin state downstream of the flipper, a \SI{6}{m} long vacuum transport system equipped with guide coils was installed, ensuring minimal depolarization during transport.

Given that interactions within the \ce{^3He} cell generate intense scattering neutron backgrounds, both the polarized \ce{^3He} and the spin flipper are positioned between ES\#1 and ES\#2 to reduce the neutron background around GTAF. Background noise from scattered neutrons and $\gamma$-rays is suppressed by collimators and the shielding doors located before ES\#2, ensuring the measurement precision of the GTAF and \ce{^6Li} neutron detectors. The main mechanical structure is made of non-magnetic aluminum alloys and brass. This minimizes magnetization from the spin flipper and guide coils and limits stray magnetic fields.

\begin{figure}[H] 
    \centering
    \includegraphics[width=0.9\linewidth]{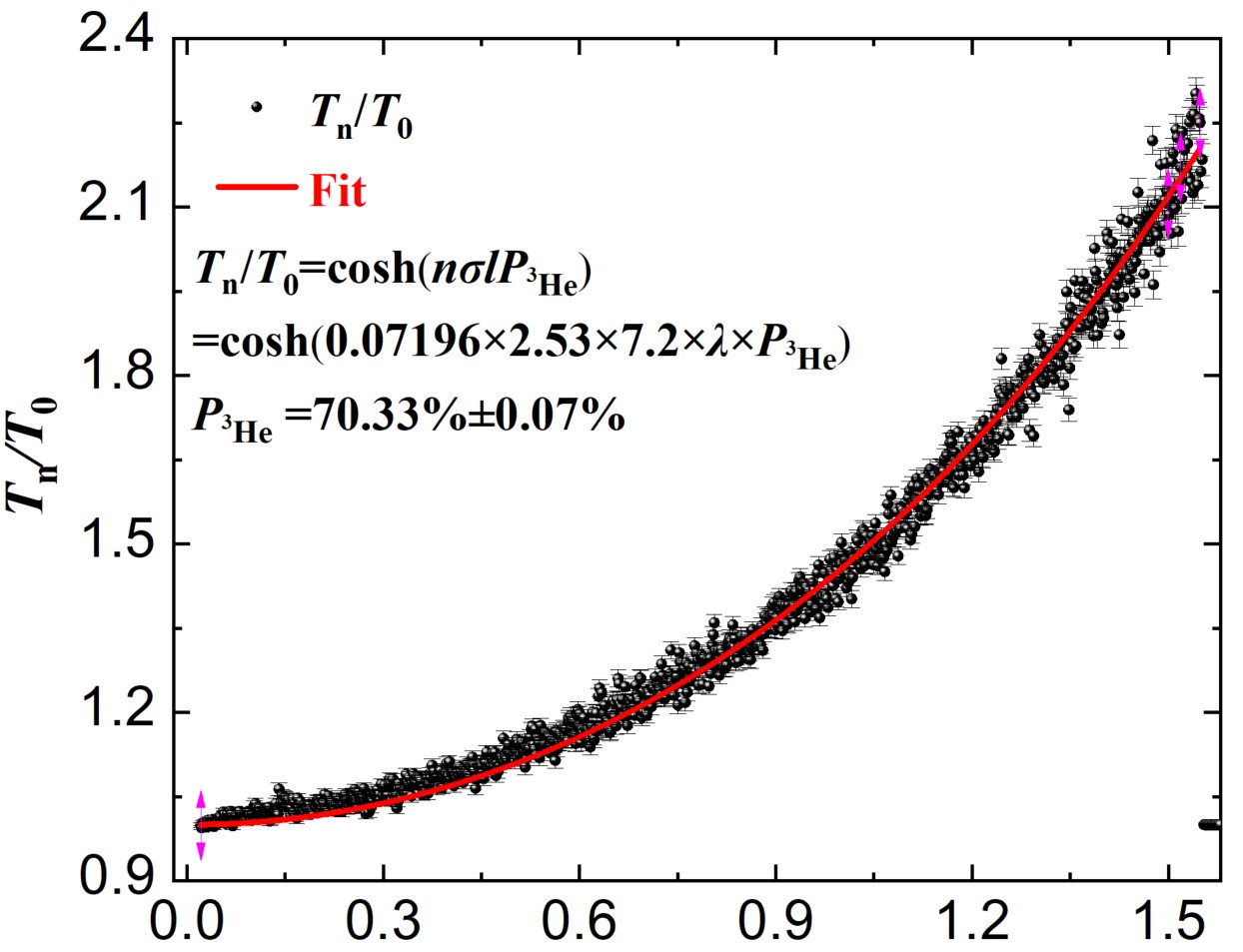} 
    \caption{Fitting result for the $^{3}\mathrm{He}$ polarization measured on Back-n}
    \label{fig:3He}
\end{figure}

\subsection{In-situ ³He system}
The in-situ \ce{^3He} system is based on Spin-Exchange Optical Pumping (SEOP) to polarize \ce{^3He} and generate polarized neutrons~\cite{zhang2022situ, qin2021development, huang2021development,zhang2025first}. Compared to other neutron polarization techniques, this method offers stable neutron polarization, polarization flipping capability, low background, a broad energy spectrum, and uniform polarization. These features are essential for achieving high-precision measurements in parity violation experiments. The system serves as the polarizer in the present setup.

Prior to the experiment, the in-situ polarized \ce{^3He} system underwent approximately several days of pre-pumping in the laboratory. The \ce{^3He} polarization at saturation was calibrated by Electron Paramagnetic Resonance (EPR), yielding a value of \SI{68}{\percent}. After confirming the system was in normal condition, the internal laser and heater were turned off, leaving only the magnetic-holding field active. The in-situ polarized \ce{^3He} device was then hoisted into the Back-n beamline, transported to the neutron beam position, and aligned. Pumping was resumed until saturation, and the system status was monitored every \SI{0.3}{h} using the Free Induction Decay (FID) method.

During the experiment, with both the adiabatic spin flipper and the GTAF detector switched off, the \ce{^6Li} detector (described in detail in Section~3.5) together with the in-situ polarized \ce{^3He} system first acquired data for two hours under saturated \ce{^3He} conditions to obtain $T_n$. After the calibration of the adiabatic spin flipper, the \ce{^3He} was depolarized, and another two-hour dataset was collected to obtain $T_0$. Because the neutron polarization is negligible at \SI{0.02}{\angstrom}, the $T_n$ and $T_0$ data were normalized using the corresponding transmission values measured at this wavelength. The \ce{^3He} polarization $P_{\ce{He}}$ is derived from the neutron transmission through the polarized \ce{^3He} Neutron Spin Filter (NSF)~\cite{ tang2025compact, zhang2022situ}:
\begin{equation}
    P_{\ce{He}} = \frac{1}{n \sigma l} \ln\left(\frac{T_n}{T_0}\right)
\end{equation}
where $T_n$ is the transmission with polarized \ce{^3He}, $T_0$ is the transmission with depolarized \ce{^3He}, $n$ is the \ce{^3He} number density, $\sigma$ is the neutron absorption cross-section, and $l$ is the cell length. The NSF used here has a \ce{^3He} pressure of \SI{2.53}{bar} and a length of \SI{7.2}{cm}. The fitted \ce{^3He} polarization was  $70.33\% \pm 0.07\%$, as shown in Fig.~\ref{fig:3He}, resulting in a neutron polarization of $30.31\% \pm  0.03\%$ at the \SI{0.734}{eV} $p$-wave resonance of Lanthanum (\ce{^{139}La}), which meets the precision requirements for parity violation measurements.

\subsection{Adiabatic spin flipper}

Precise TRIV and PV measurements depend on determining transmission asymmetries, necessitating stable spin manipulation at epithermal energies (specifically the \SI{0.734}{eV} resonance and above). Conventional Radio Frequency (RF) flippers are impractical in this regime, as satisfying the adiabatic condition for fast neutrons requires excessive RF power and elongated gradient fields. Such high-power operation inevitably generates electromagnetic radiation, imposing severe shielding requirements and risking electronic circuit instability. Consequently, standard RF designs are unsuitable, necessitating an alternative approach.

Beyond engineering constraints, detector stability presents a further challenge. Long-duration experiments are susceptible to temperature-induced detector gain drift, which can introduce false asymmetries. To mitigate this, spin states must be switched frequently to cancel thermal drift~\cite{schaper2020modular}. The polarized \ce{^3He} system alone is insufficient, as it cannot reverse spin direction at the required high frequency (e.g., every few seconds) to effectively suppress these errors.

To meet these NOPTREX requirements, an adiabatic spin flipper was designed downstream of the polarizer. Distinct from RF methods, this device achieves adiabatic spin reversal through a spatially rotating magnetic field in the ON state, while introducing a magnetic field zero-crossing in the OFF state to permit polarization transmission. This independent configuration allows for rapid spin switching (e.g., every \SI{0.4}{s}) to cancel detector drift without reversing the \ce{^3He} polarization. Additionally, specific switching cycles are employed to eliminate systematic errors caused by magnetic field asymmetries.

\begin{figure}[H] 
    \centering
    \includegraphics[width=0.9\linewidth]{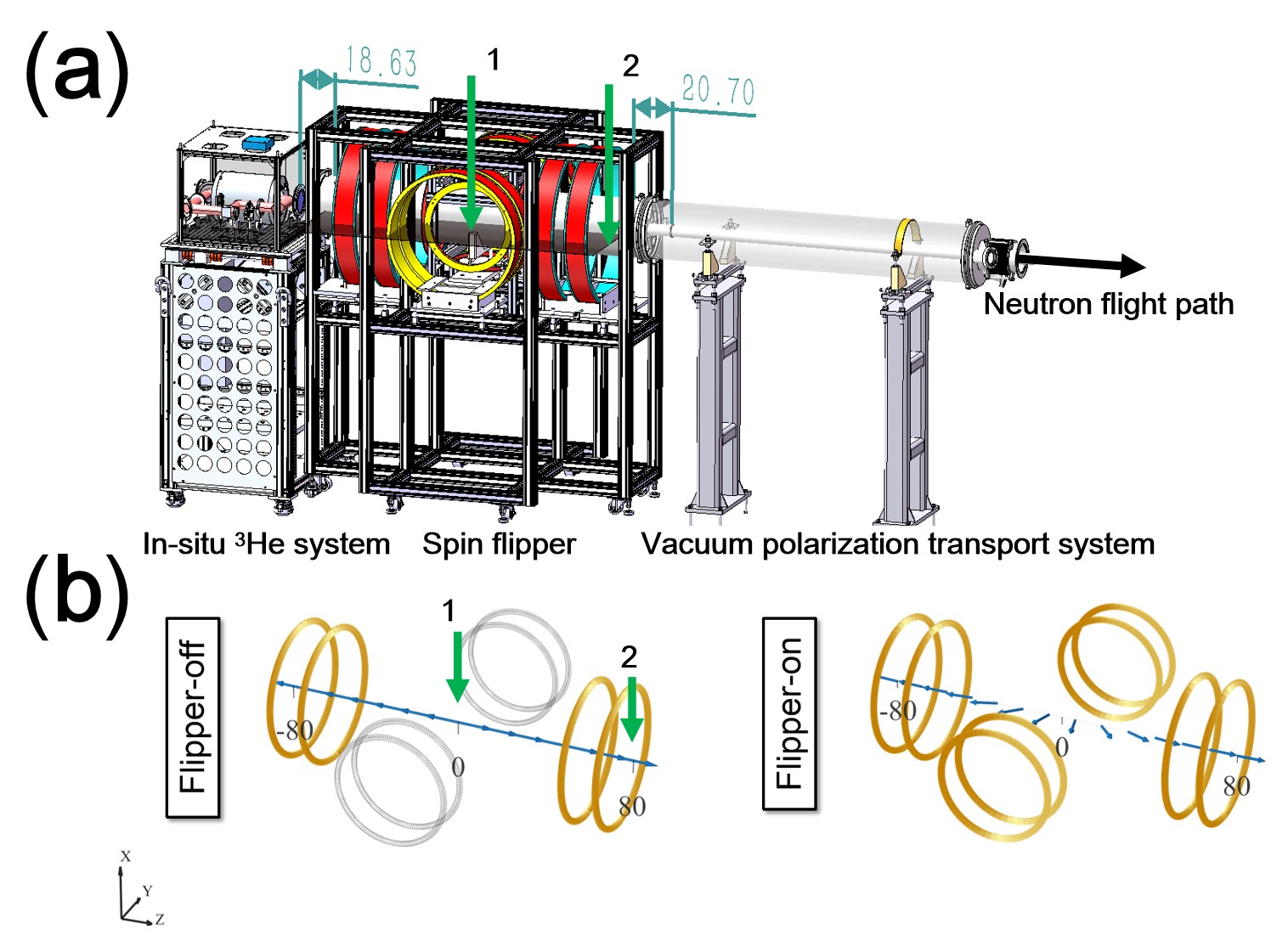} 
    \caption{\textbf{(a)} Mechanical assembly of the part polarization, comprising the \textit{in-situ} \ce{^3He} system, the spin flipper, and the vacuum polarization transport system. The entrance and exit of the flipper are defined at \SI{80}{cm} from its geometric center. The labels 1 and 2 denote the geometric center ($z=\SI{0}{cm}$) and the exit plane ($z=\SI{80}{cm}$) of the adiabatic spin flipper, respectively. The axial separation between the \ce{^3He} system and the flipper entrance is \SI{18.63}{cm}, while the distance from the flipper exit to the transport system is \SI{20.7}{cm}. \textbf{(b)} Spatial distribution of the magnetic field vectors, illustrating the field topology in both the flipper-on and flipper-off states.}
    \label{fig:Mechanical assembly}
\end{figure}

\subsubsection{Physical field design}

The spin flipper, the upstream in-situ \ce{^3He} polarizer, and the downstream polarization transport system are all located in End Station~2 (ES\#2). Due to the spatial constraints of the spectrometer line, the flipper was constructed with a length of \SI{1.75}{m} and a width of \SI{1.27}{m}, leaving a clearance of less than \SI{10}{cm} from the side wall. The upstream polarizer features compact dimensions---\SI{48}{cm} in height, \SI{55}{cm} in length along the beam, and \SI{56}{cm} in width---with a magnetic field confined strictly to its geometric volume. The downstream magnetic guide consists of a \SI{6}{cm} diameter solenoid with a field extending \SI{3}{cm} outward. As shown in Fig.~\ref{fig:Mechanical assembly}(a), based on this geometry, the device is integrated into the beamline with an \SI{18.63}{cm} gap separating the entrance from the polarizer, and a \SI{20.7}{cm} gap between the exit and the downstream magnetic guide field system, ensuring seamless polarization transport.

\begin{figure}[H] 
    \centering
    \includegraphics[width=0.9\linewidth]{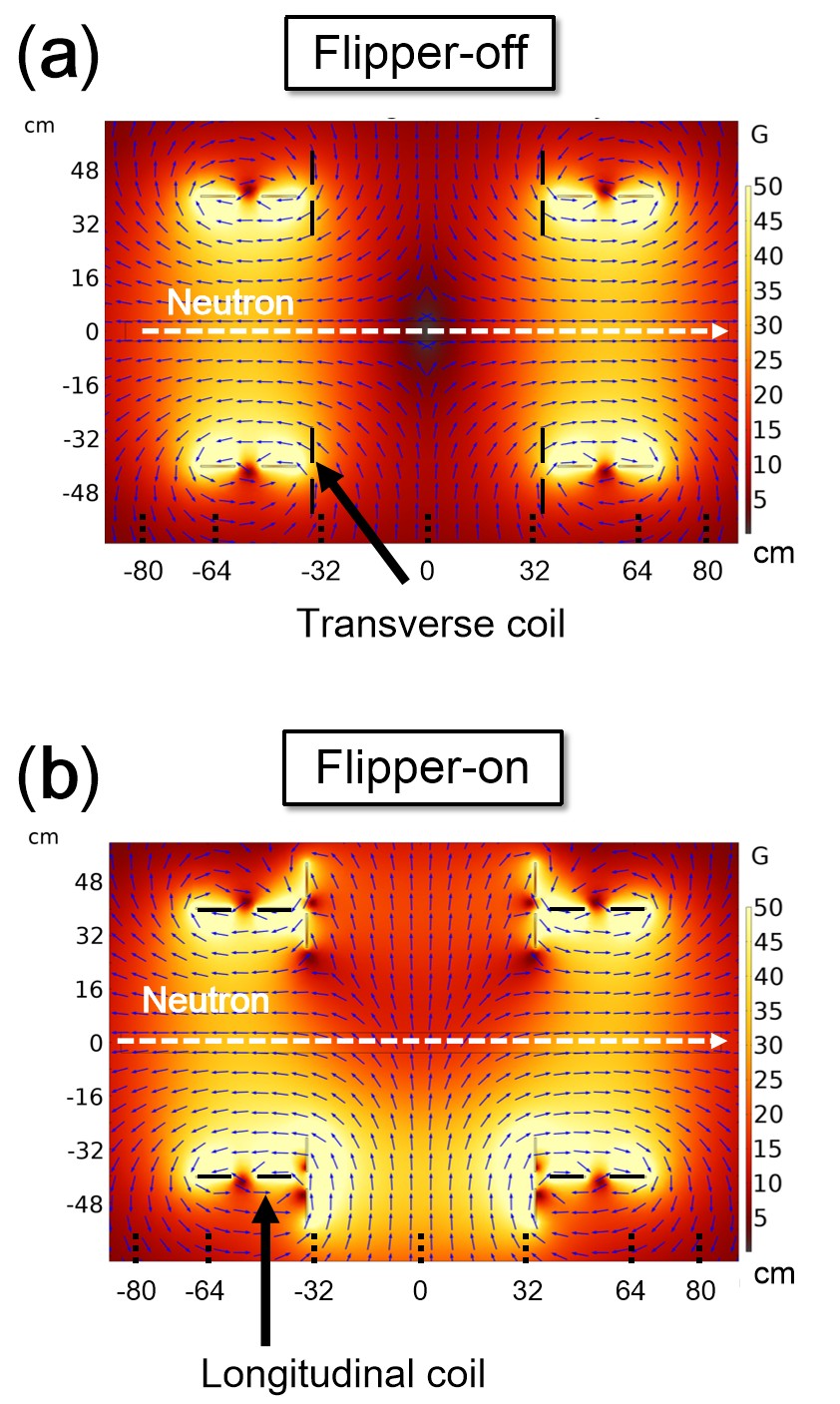} 
    \caption{\textbf{(a)} and \textbf{(b)} show the magnetic field distribution on the cross-section of the mechanical structure depicted in Fig.~\ref{fig:Mechanical assembly}(a), corresponding to the spin flipper being in the flipper-on and flipper-off states, respectively.}
    \label{fig:field_distribution}
\end{figure}

As illustrated in Fig.~\ref{fig:Mechanical assembly}(a), a custom coil group configuration was implemented to optimize adiabatic spin-flip efficiency for eV neutrons. Two symmetric coil groups ($R=\SI{40}{cm}$) separated by \SI{80}{cm} generate the longitudinal gradient; this geometry widens the transition zone to satisfy adiabatic conditions while maintaining field homogeneity. The spin flip is driven by four transverse coils ($R=\SI{36}{cm}$) with a \SI{57}{cm} separation, which provide the rotating field vector along the neutron path. Field mapping confirms that the longitudinal zero-crossing point deviates by less than \SI{2}{mm} from the geometric center. This alignment demonstrates the consistency between the mechanical assembly and the magnetic design, thereby validating the use of magnetic simulations to characterize the device performance. Fig.~\ref{fig:Mechanical assembly}(b) illustrates the variation of the magnetic field vectors along the neutron path for both the flipper-off and flipper-on states. Mechanically, the support structure is fabricated entirely from non-magnetic aluminum alloy to eliminate hysteresis-induced stray fields. The assembly is aligned using an external laser positioning system, constraining the deviation between the magnetic coil axis and the neutron beam center to within \SI{1}{mm}.

To optimize the magnetic field distribution for eV neutron spin manipulation, 3D finite element simulations were performed using the AC/DC module of COMSOL Multiphysics~\cite{comsol}. The model incorporated coil geometries, defining the longitudinal and transverse coils with uniform multi-turn windings carrying excitation currents of \SI{16}{A} and \SI{10}{A}, respectively. This high-fidelity FEM model was rigorously validated against experimental 3D field mapping data and served as the primary framework for analyzing neutron polarization evolution. Crucially, the simulation reveals that the longitudinal fringe fields extend sufficiently to bridge the \SI{10}{cm} gaps separating the flipper from the upstream in-situ \ce{^3He} system and the downstream transport system. This confirms that the flipper provides the continuous guiding field necessary to preserve polarization, enabling effective characterization of spin transport dynamics without the computational overhead of modeling the localized internal fields of adjacent components.

Fig.~\ref{fig:field_distribution}(a) and (b) show the calculated field distribution in the flipper-on and flipper-off states, respectively. Compared with the flipper-on state, the flipper-off configuration produces a field minimum (``null magnetic field'') near the geometric center of the system, as highlighted in Fig.~\ref{fig:field_distribution}(b). However, complex ambient magnetic fields within the spectrometer can induce a spatial drift of this zero point. Such displacement risks compromising both the polarization flipper-on and flipper-off efficiencies, ultimately limiting the measurement precision of the NOPTREX experiment. To mitigate this, three sets of compensation coils were installed to nullify residual stray field components along the $x$, $y$, and $z$ axes at the geometric center, stabilizing the flipper's performance. Each coil set features a diameter of \SI{23}{cm} and is positioned symmetrically at a distance of \SI{47.5}{cm} from the flipper's geometric center. In the flipper-on state, the longitudinal profile remains unaltered, while a transverse ($y$-axis) component with a sinusoidal spatial dependence is superimposed. This transverse field vanishes at both the entrance and exit, peaking at \SI{16}{G} at the center.

\subsubsection{Neutron spin behavior in magnetic Fields}
\begin{figure}[H] 
    \centering
    \includegraphics[width=0.9\linewidth]{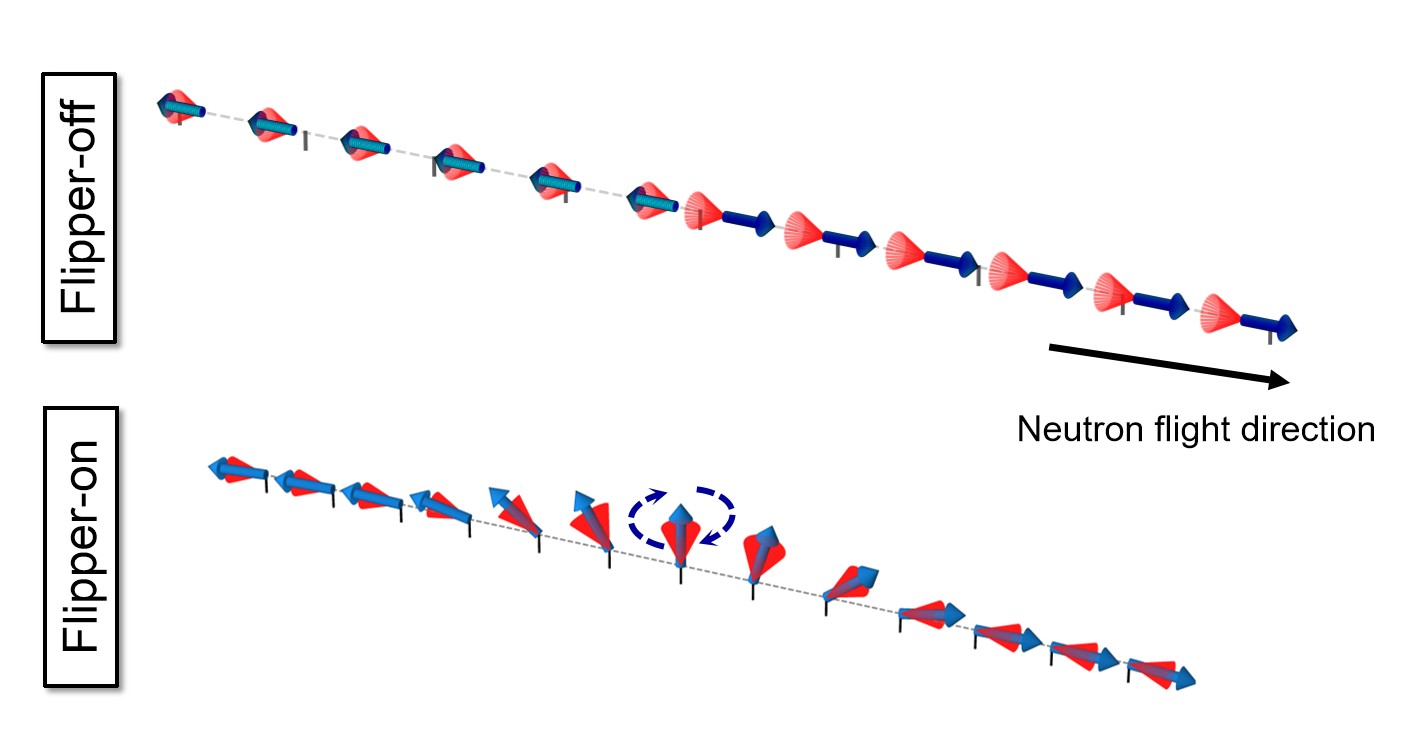}
    \caption{Schematic illustration of the wavelength of \SI{0.34}{\angstrom} neutron spin precession in the flipper-on and flipper-off states. The view is magnified to visualize the precession cone corresponding to a \SI{10}{\degree} misalignment between the spin vector and the magnetic field. The blue arrows indicate the direction of the local magnetic field vectors ($\bm{B}$). The red conical surfaces represent the precession trajectories of the spins around the magnetic field, where the cone opening angle corresponds to the precession angle ($\theta$).}
    \label{fig:precession_schematic}
\end{figure}

To enable efficient and stable spin manipulation for measurements requiring polarized neutrons in the eV energy range, we adopt the adiabatic spin flipper developed by Roberson et al.~\cite{roberson1993apparatus,schaper2020modular, bowman1996spin}. The device generates a superposition of longitudinal and transverse magnetic fields to flip the neutron spin via the adiabatic condition. The neutron spin precession is governed by the Larmor frequency $\omega_L = \gamma_n B$, where $\gamma_n$ denotes the neutron gyromagnetic ratio and $B$ represents the magnetic field. In the laboratory frame, the rotation frequency of the magnetic field is expressed as $\omega_B = v/L$, where $v$ is the neutron velocity and $L$ characterizes the length of the field scale. The adiabatic condition requires that the magnetic field direction changes much slower than the precession rate, quantified by the adiabatic parameter $k=\omega_L/\omega_B$. When $k \gg 1$, the neutron spin follows the slowly rotating magnetic field, resulting in a spin flip.

If the flipper is off (transport mode), the transverse magnetic field vanishes, while the longitudinal field reverses polarity at the geometric center. When polarized neutrons traverse this zero-crossing point, those satisfying the adiabatic condition ($k>1$) will adiabatically track the field reversal, resulting in an unintended spin flip. Consequently, the device exhibits polarization loss for low-energy neutrons ($k>1$). In contrast, high-energy neutrons ($k<1$) undergo non-adiabatic transport through the zero-field region, preserving their spin state and ensuring flipper-off efficiency.

\subsubsection{Simulation of flipper efficiency}
To verify the magnetic-field distribution used for neutron spin manipulation at eV energies in NOPTREX, we utilized a numerical algorithm based on the Bloch equations~\cite{seeger2001numerical}. This method employs a sixth-order Runge-Kutta integration scheme for the calculation of the neutron polarization vector's evolution under complex magnetic field conditions, facilitating the assessment of both flipper-on and flipper-off efficiencies. The model resolves the spatial evolution of critical parameters, including the neutron polarization vector and the adiabatic parameter. For initialization, the incident neutron polarization vector is assumed to be aligned parallel to the local magnetic field at the entrance of the spin flipper.

Following this initialization, we tracked the spatial evolution of the angle $\theta_{P,B}$ between the polarization vector and the magnetic field vector. In Fig.~\ref{fig:precession_schematic}, for a wavelength of \SI{0.34}{\angstrom}, the neutron spin vector maintains a tight parallel alignment with the magnetic field upon entering the flipper, exhibiting a minimal angular deviation of $\sim \SI{1}{\degree}$. As the magnetic field vector begins to rotate, the adiabaticity parameter decreases due to the weakening effective field. Consequently, the angle $\theta_{P,B}$ gradually expands from \SI{1}{\degree} to a maximum of $\sim \SI{10}{\degree}$. Despite this transient divergence, the spin vector effectively tracks the evolving field direction. Finally, as the neutron exits the flipping region, the spin-field alignment is restored, with $\theta_{P,B}$ relaxing back to approximately \SI{3}{\degree}. Given the experimental requirements, these residual oscillations are sufficiently small and are expected to have a negligible impact on the overall precision of the NOPTREX measurements.

\begin{figure}[H] 
    \centering
    \includegraphics[width=0.9\linewidth]{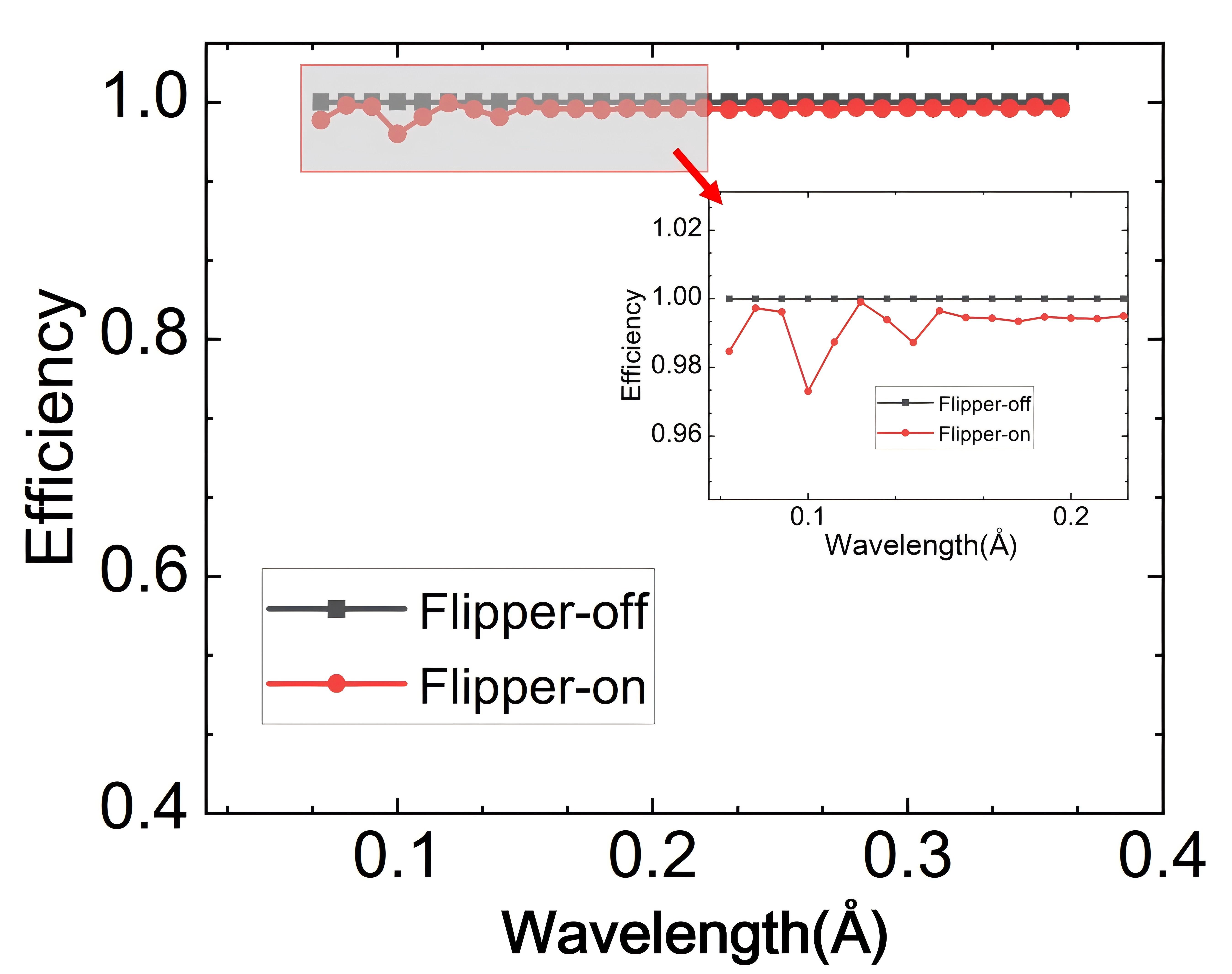}
    \caption{The flipper-off and flipper-on efficiencies calculated from the two magnetic field distributions of the spin flipper presented in Fig.~\ref{fig:field_distribution}.}
    \label{fig:efficiencies}
\end{figure}

To demonstrate the capability of this magnetic field configuration to meet the prospective requirements of NOPTREX, specifically for parity violation measurements in the epithermal energy range, we analyzed the multi-wavelength flipper-on efficiency as a function of the flipper magnetic field. Fig.~\ref{fig:efficiencies} presents the flipper-off (black dots) and flipper-on efficiencies (red dots) calculated for neutron wavelengths ranging from \SI{0.09}{\angstrom} to \SI{0.36}{\angstrom}. The flipper-off data indicate that for neutrons traversing the zero-field point, the spin state is unaffected, nearly preserving the initial polarization vector. The flipper-on efficiency exhibits persistent oscillations with respect to neutron wavelength, characterized by a diminishing amplitude as the wavelength increases. For wavelengths exceeding \SI{0.2}{\angstrom}, the efficiency stabilizes above \SI{97}{\percent}, approaching saturation. This phenomenon is intrinsic to the physics of adiabatic spin flippers ~\cite{grigoriev1997peculiarities}. For an ideal case where the maximum transverse and longitudinal field amplitudes are identical and the field distribution is sinusoidal, the relationship between the spin flipping efficiency and the adiabaticity parameter is characterized by periodic oscillations. Although the magnetic field distribution in the design deviates slightly from the ideal sinusoidal profile, the observed oscillatory trend closely follows the theoretical prediction described by this formula. These simulation results demonstrate that the magnetic field of the spin flipper is capable of manipulating neutron spins across the \SIrange{0.7}{10}{eV} energy range, satisfying the requirements for NOPTREX PV measurements.

Following the design by Roberson~\cite{roberson1993apparatus}, the switching sequence alternates between `N' (flipper-off) and `F' (flipper-on) states. This sequence is designed to cancel the effects of first-order transverse coil stray fields and eliminate both linear and quadratic time drifts in the detector. The polarity of the transverse magnetic field is defined by the current direction: a clockwise current ($+$) generates a field parallel to the $+y$-axis, while a counter-clockwise current ($-$) generates a field parallel to the $-y$-axis. Based on this control logic, the operational sequence for one complete cycle of the spin flipper is defined as: $+$, $0$, $0$, $-$, $0$, $-$, $+$, $0$. For this modulation scheme to be effective, strict temporal synchronization between the magnetic state transitions and the data acquisition system is mandatory. This ensures that measurements corresponding to distinct spin states are correctly isolated, allowing for the accurate determination of asymmetry values required by NOPTREX.

The measurement of PV relies on detecting differences in scattering signals between distinct spin states. Therefore, it is essential to synchronize data acquisition with the operational state of the spin flipper. To achieve this goal, a dedicated signal feedback system was integrated into the flipper circuit. This feedback signal tracks the output current of the transverse coil in real time, with its magnitude accurately reflecting the flipper's state. By utilizing the spallation source's $t_0$ pulse as a common time reference, the detector data is tagged with the flipper feedback signal. This data acquisition system maintains a synchronization error of less than \SI{1}{ms}, ensuring the precision required for PV measurements in the NOPTREX experiment.

\subsection{Vacuum polarization transport system}
To suppress scattering signals generated by epithermal energy neutrons interacting with the in-situ \ce{^3He} system, the polarizer and spin flipper were positioned between ES\#1 and ES\#2, utilizing collimators and shielding walls to minimize signal interference. The spin flipper exit is situated \SI{7.5}{m} upstream of GTAF ~\cite{qi2021cross}. Gaussmeter measurements confirm that the ambient magnetic environment is dominated by the geomagnetic field, with a strength of approximately \SI{0.4}{G}. Over this flight path, the uncompensated field is sufficient to induce a precession angle exceeding \SI{100}{\degree} for \SI{0.74}{eV} neutrons.

To preserve the neutron polarization after the spin flipper and minimize air scattering, a vacuum polarization transport system was designed and installed. The layout of this system is illustrated in Fig.~\ref{fig:schematic}, which details the guide field distribution within the eV polarized neutron. The guide field begins \SI{13}{cm} downstream of the spin flipper exit, where the magnetic field maintains a minimum intensity of \SI{12}{G} at the interface.  Given the requisite guide field length of approximately \SI{6}{m}, the transmission system employs a segmented design consisting of four interconnected solenoids. The first three solenoids are installed within the vacuum chamber, where spatial constraints limit their maximum outer diameter to \SI{6}{cm}. The fourth solenoid, positioned outside the vacuum section, features a larger diameter of \SI{16}{cm}. To ensure precise beam alignment, collimation supports are installed for both the internal and external sections. Consequently, small axial gaps exist at these support structures and between the individual solenoids. The transport line extends approximately \SI{6.2}{m} to the front face of the GTAF, passing through a collimator section. The collimator is equipped with an internal guide field. There is a \SI{3}{cm} axial gap separating the upstream guide solenoid from this internal collimator field. At this interval, the magnetic field reaches a local minimum of \SI{14}{G}. The guide field solenoid is wound with three layers of \SI{1}{mm} diameter copper wire. The multi-layer copper winding configuration minimizes current density and heat generation, stabilizing the magnetic environment under vacuum conditions. A minimum aperture of \SI{56}{mm} within the collimator ensures ample clearance for the \SI{30}{mm} beam. At an operating current of \SI{0.8}{A}, the system sustains a guide field minimum of \SI{12}{G}, adequate for the spin transport requirements of NOPTREX.

\subsection{Detector Polarization Preservation}
To characterize the performance of the experimental equipment (such as the spin flippers during double-flip tests), neutron counting was performed using a \ce{^6Li}-based detector~\cite{ref32}. This detector utilizes Cerium-activated lithium glass (\ce{Ce}-doped \ce{Li2O \cdot 2SiO2(Ce)}), a favored material for neutron instrumentation due to its physicochemical robustness and resilience across a broad temperature range. The specific assembly incorporates glass enriched to \SI{95}{\percent} \ce{^6Li} to ensure sensitivity to thermal neutrons. The scintillator element consists of a \SI{5}{cm} $\times$ \SI{5}{cm} $\times$ \SI{0.1}{cm} crystal coupled to a photomultiplier tube (PMT) with a \SI{34}{mm} diameter window and a peak spectral response at \SI{420}{nm}. At a neutron energy of \SI{0.734}{eV}, this detector demonstrates an efficiency of \SI{25}{\percent}, which is sufficient for the calibration requirements of this experiment.

The primary objective of the NOPTREX experiment is the precise determination of PV and TRIV. Since NOPTREX measurements based on the $(n, \gamma)$ reaction offer a sensitivity enhancement of two orders of magnitude compared to neutron transmission experiments, the GTAF detectors are combined with the eV polarized neutron system to form the CSNS NOPTREX measurement setup. The GTAF assembly comprises 40 \ce{BaF2} crystals (12 pentagonal, 28 hexagonal) arranged in a spherical shell geometry (inner radius \SI{10}{cm}, thickness \SI{15}{cm}) that covers \SI{95.2}{\percent} of the solid angle~\cite{qi2021cross}. Individual crystals ($\varnothing$\SI{14}{cm} $\times$ \SI{15}{cm}) are wrapped in \SI{2}{\micro\meter} Teflon and \SI{1}{\micro\meter} aluminum foil, secured with light-tight black tape, and optically coupled to PMTs using silicone grease within aluminum housings. This configuration yields an energy resolution of \SI{10}{\percent} at \SI{1274}{keV} and a time resolution of \SI{677 \pm 32}{ps}, enabling precise incident neutron energy reconstruction via time-of-flight detection of capture gamma cascades.

Regarding the beam transport immediately upstream of the GTAF, there is a \SI{1.2}{m} section that lacks a magnetic guide field, though a vacuum beam pipe is installed to minimize air scattering. While this gap interrupts the continuous polarization transport, verification experiments have confirmed that the resulting effect on polarization, although present, does not compromise the validity of the final experimental results. A detailed quantitative analysis of the depolarization effects induced by these guide field discontinuities, along with corresponding simulation results, is presented in Section 5.

\section{Experimental Results}\label{sec:4}
\subsection{Measurement of the asymmetry of \ce{^{139}La}}
\begin{figure}[H] 
    \centering
    \includegraphics[width=0.9\linewidth]{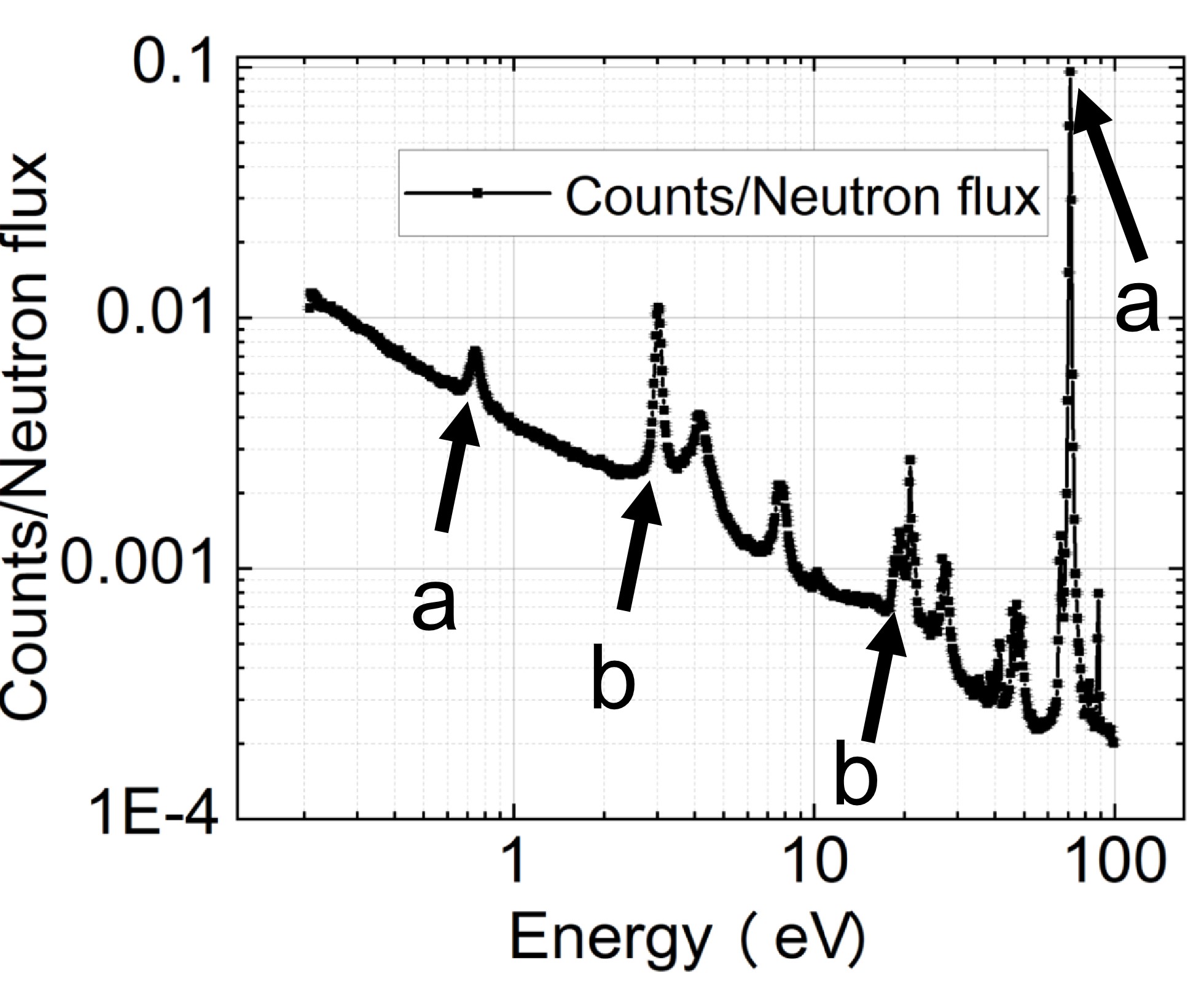} 
    \caption{Energy dependence of the $\gamma$-ray yields normalized by the incident neutron flux. The arrows labeled \textbf{(a)} indicate the \ce{^139La} resonance peaks at \SI{0.74}{eV} and \SI{71}{eV}, while \textbf{(b)} marks the \ce{^138La} resonances located at \SI{3.0}{eV} and \SI{21}{eV}.}
    \label{fig:energy_dependence}
\end{figure}

A room-temperature \ce{^{139}La} disk with a radius of \SI{40}{\mm} and a thickness of \SI{1.8}{\mm} was aligned orthogonal to the neutron beam axis at the center of the GTAF. Fig.~\ref{fig:energy_dependence} presents the measured neutron capture yield for the natural lanthanum target. Background suppression was implemented by requiring a $\gamma$-ray multiplicity of $M \ge 3$ and a total energy deposition of $E_{\gamma} > \SI{3}{MeV}$. In the off-resonance regions, the capture yield follows the characteristic $1/v$ dependence, manifesting as a linear trend on a double-logarithmic scale. This adherence to the expected cross-section behavior confirms that residual backgrounds are negligible (systematic errors of the asymmetry measurement are discussed late)~\cite{endo2023measurements}.

The $p$-wave resonance of \ce{^{139}La} at \SI{0.74}{eV} is clearly resolved from the neighboring peaks at \SI{3}{eV} and \SI{21}{eV}, which are identified from the \ce{^{138}La} isotope. This spectral separation is a prerequisite for the parity violation measurement. The detection capability extends to higher energies, clearly resolving the strong \ce{^{139}La} $s$-wave resonance at \SI{71}{eV}. Although the use of an uncalibrated natural target implies the potential presence of trace rare-earth contaminants, the dominant spectral features align with the known cross-sections of natural lanthanum.

\begin{figure}[H] 
    \centering
    \includegraphics[width=0.9\linewidth]{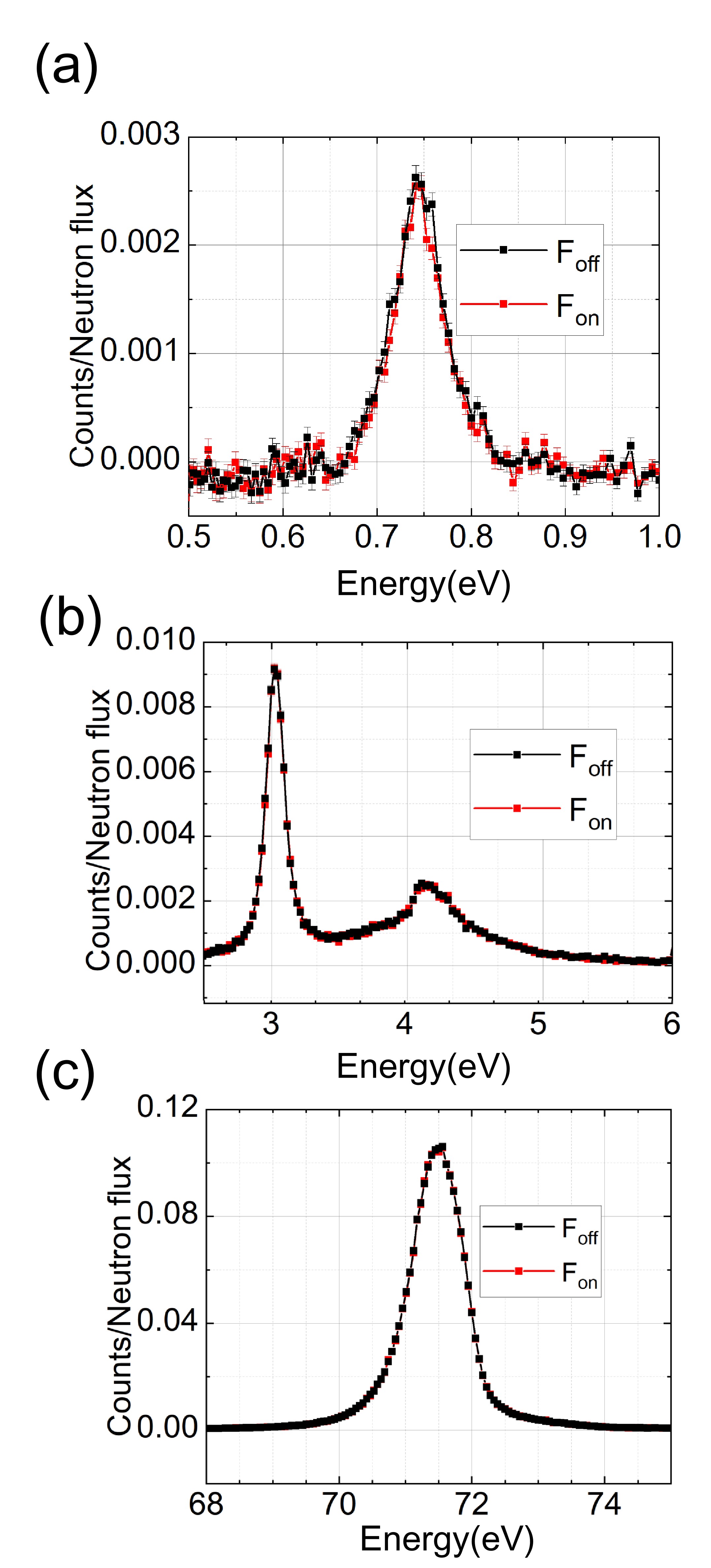} 
    \caption{ (a) The asymmetry of La at \SI{0.734}{eV} measured by the GTAF $\gamma$ detector, derived from the spin flipper-on/flipper-off states. (b) The asymmetry calculated at \SI{3}{eV}, yielding a value of $0.37\% \pm 0.31\%$. (c) The asymmetry at \SI{71}{eV}, yielding $0.05\% \pm 0.34\%$.}
    \label{fig:asymmetry_la}
\end{figure}

Following the confirmation that the GTAF signals were free from significant background interference, PV measurements were conducted across different spin states. The spin flipper operated according to the previously described eight-step sequence, switching states every 10 neutron pulses with a total cycle period of \SI{3.2}{s}. Data acquisition was synchronized with the spin flipper feedback to correlate each GTAF event with its corresponding polarization state. The experimental asymmetry was derived via the expression $ Asymmetry = (N_{\text{trans}} - N_{\text{flip}}) / (N_{\text{trans}} + N_{\text{flip}})$, where $N$ represents the resonance peak area normalized by the incident neutron flux. For the \SI{0.747}{eV} resonance shown in Fig.~\ref{fig:asymmetry_la}, this analysis yields an asymmetry of \num{4.67 \pm 1.19}\,\si{\percent}. The total uncertainty is dominated by statistical contributions arising from two primary sources: the counting statistics of the detected $(n, \gamma)$ events, and the statistical fluctuations in the incident neutron flux measured by the Li-glass detector. These uncertainties were rigorously propagated through the peak integration and asymmetry calculations. 

The asymmetry at the $70~\text{eV}$ resonance is consistent with zero within errors $0.05\% \pm 0.34\%$. This is a good check on many possible sources of systematic error in the apparatus, such as normalization errors, electronic cross-talk in the data acquisition system correlated with the neutron spin flipper activation, and possible magnetic field effects of the neutron spin flipper on the gamma detectors. At $70~\text{eV}$, however, the neutron beam polarization is nearly zero. It is therefore also valuable to analyze the asymmetry on neutron resonances that are closer in energy to the $0.7~\text{eV}$ $p$-wave of interest. In our sample, we observe resonance signals at $3.0~\text{eV}$ and $4.0~\text{eV}$ which we can use for this purpose. The $3.0~\text{eV}$ resonance is a known $s$-wave resonance in $^{138}\text{La}$, which is present in the sample at low abundance. The $4.0~\text{eV}$ resonance we observe is not from La and likely originates from an impurity nucleus in the sample, it is also an $s$-wave resonance. Therefore, in view of the very small amplification of parity violation in $s$-wave resonances compared with $p$-wave resonances predicted by theory, we would expect an asymmetry of zero within errors at our level of measurement precision. We therefore constructed an integral asymmetry from $2.5~\text{eV}$ to $6~\text{eV}$, which contains both the $3~\text{eV}$ and $4~\text{eV}$ resonances, after subtracting the $1/v$ potential scattering background based on the fit to the $0.74~\text{eV}$ resonance. Using this integral value, we calculated a raw asymmetry of $0.37\% \pm 0.31\%$. This asymmetry is also consistent with zero, but in this case, the neutron polarization is much larger than at $70~\text{eV}$ ($polarization = 15.66\% \pm 0.01\%$ at $2.5~\text{eV}$ and $polarization = 10.22\% \pm 0.01\%$ at $6~\text{eV}$). So in this case, the gamma detectors see a signal that is coming from polarized neutrons absorbing in the La target. The absence of an asymmetry shows that there are no obvious systematic errors connected with polarized neutron interactions in the target material (this could in principle come from magnetic impurities in the sample, for example). It is also an additional check on possible systematic errors in normalization, electronic cross-talk, magnetic field effects, etc., simply because the events appear in the detector at a different neutron time of flight arrival. Taken together, the zero asymmetries from the analysis of the resonance signals at these two different neutron energies, different neutron times of flight, and different neutron polarization values at the target are strong evidence for the absence of systematic errors at the $0.7~\text{eV}$ resonance at this level of precision.

\subsection{Measurement about the spin flipper efficiency}
Raw spectra acquired by the \ce{^6Li} neutron detector were analyzed to extract physical observables by determining and subtracting the background contribution. This background originates primarily from proton-induced target $\gamma$-rays and neutrons undergoing multiple reflections within the neutron guide. The profile was characterized using the black resonance filter technique with \ce{Cd}, \ce{Ag}, and \ce{Co} absorbers. By fitting the residual counts at the saturated resonance minima using the parameterization described in Ref.~\cite{schillebeeckx2012determination}, the time-correlated background components were effectively isolated and removed. This precise background correction is a prerequisite for accurately calculating the spin flipper efficiency, a critical system parameter for the asymmetry measurements discussed below.

The capability of the eV polarized neutron beamline for asymmetry measurements was validated via the \ce{^{139}La} $p$-wave resonance at \SI{0.74}{eV}. Since the polarization transport system operates with a static guide field, its efficiency is convolved with that of the spin flipper and is not characterized independently. Unlike standard resonant flippers, the adiabatic spin flipper exhibits non-unity efficiencies in both the flipper-on and flipper-off states. However, accurate extraction of the PV requires calibration of the spin flipper efficiency at the target energy. This calibration necessitates a double-flip measurement, requiring an expansion of the instrumentation. To complement the existing in-situ \ce{^3He} polarizer (detailed in Section II; \SI{30}{\percent} polarization for \SI{0.74}{eV} neutrons), the beamline was instrumented with an off-situ \ce{^3He} analyzer and the aforementioned \ce{^6Li} neutron detector. The in-house developed in-situ pumped \ce{^3He} spin filter employs adiabatic fast passage (AFP) to manipulate the \ce{^3He} spin states, functioning as a neutron analyzer~\cite{zhang2022situ}. On-beam operation confirms stable performance with a lifetime exceeding \SI{140}{h} and a maximum saturated \ce{^3He} polarization of \SI{67}{\percent}, fully satisfying the requirements for flipper efficiency calibration.

The flipper-on efficiency was evaluated using a double-flip scheme comprising four neutron transmission measurements. This method permutes the state of the spin flipper (flipper-on/flipper-off) and the orientation of the \ce{^3He} NSF polarization relative to the magnetic field (parallel/anti-parallel)~\cite{jones1978non, williams1988polarized}. We designate the efficiencies of the in-situ \ce{^3He} polarizer, off-situ \ce{^3He} analyzer, and spin flipper as $p$, $a$, and $f$, respectively.

The spin transport through the optical line is modeled using a transfer matrix formalism. Assuming an unpolarized incident beam with unit flux, represented by the spinor $N_0 = (0.5, 0.5)^\mathrm{T}$, the system is described by the operator sequence $AFP$. The polarization and analysis operators are defined as diagonal matrices $P = \operatorname{diag}((1+p)/2, (1-p)/2)$ and $A = \operatorname{diag}((1+a)/2, (1-a)/2)$.
. The spin flipper is represented by a symmetric transfer matrix $F$, which accounts for the spin-state efficiency $\varepsilon$:
\begin{equation}
M = \begin{pmatrix}
\frac{1+\varepsilon}{2} & \frac{1-\varepsilon}{2} \\
\frac{1-\varepsilon}{2} & \frac{1+\varepsilon}{2}
\end{pmatrix}
\end{equation}

The transmitted intensity is obtained by summing the components of the final spinor $N_{\text{out}} = A M P N_0$. Incorporating the transmission factors for the unpolarized beam ($T_0$), polarizer ($T_f$), analyzer ($T_a$), and the flipper in the flipper-off state ($T_{\text{trans},\varepsilon}$), the total intensity for the parallel ($a$) and anti-parallel ($-a$) configurations in the transport mode is derived as:
\begin{equation} 
M_{1,2} = \frac{1}{4} (1 \pm a \varepsilon p) T_0 T_a T_{\text{trans},\varepsilon} T_p. 
\end{equation} 
Here, the positive and negative signs correspond to the parallel and anti-parallel analyzer orientations, respectively. It is important to note that for the flipper-on state, the efficiency parameter undergoes a sign reversal ($\varepsilon \rightarrow -\varepsilon$) due to the permutation of the matrix elements. Measurements M3 and M4 replicated these analyzer orientations with the flipper-on state.

\begin{figure}[H] 
    \centering
    \includegraphics[width=0.9\linewidth]{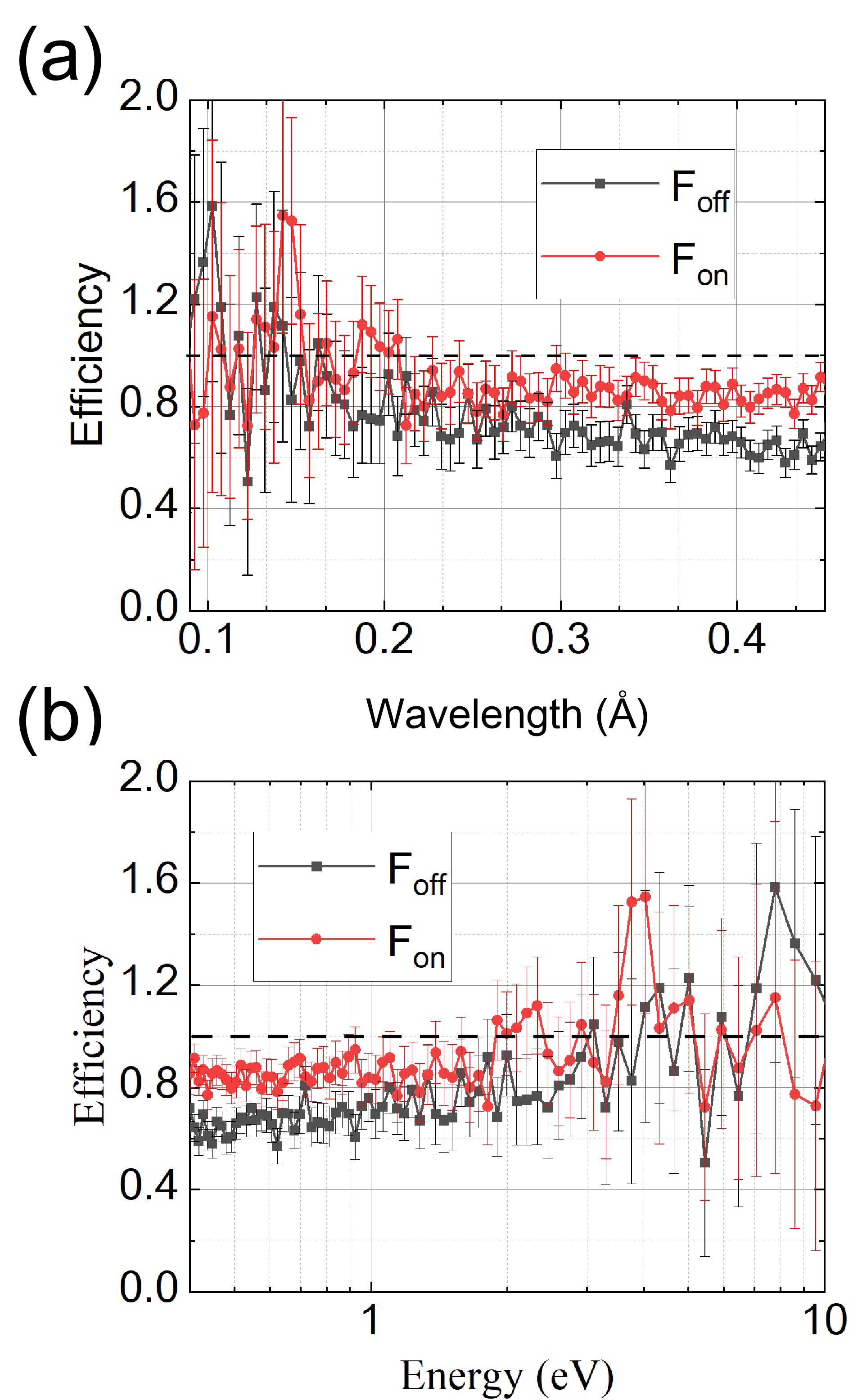} 
    \caption{The polarization flipper-off efficiency and flipper-on efficiency measured at different wavelengths (energies) in the experiment.}
    \label{fig:efficiency_wavelength}
\end{figure}

To characterize the spin modulation capability, we refer to the standard definition of flipper
-on efficiency, $\varepsilon$, where the output polarization $f_{\text{out}}$ is related to the incident polarization $f_{\text{in}}$ by the scalar relationship $f_{\text{out}} = f_{\text{in}}(1-2\varepsilon)$. However, unlike conventional designs, the spin flipper utilized in this study does not achieve ideal operation in either the flipper-on or the flipper-off state. Consequently, we define distinct efficiencies for each regime. The flipper-on efficiency ($F_{\text{on}}$) aligns with the traditional definition, governing the polarization reversal as $f_{\text{out}} = f_{\text{in}}(1-2\varepsilon_{\text{on}})$. In contrast, the flipper-off efficiency ($F_{\text{off}}$) is defined to quantify the preservation of the spin state, described by the relation $f_{\text{out}} = f_{\text{in}}(2\varepsilon_{\text{off}}-1)$. Under these definitions, a value of unity for either efficiency corresponds to the ideal performance for that specific operational mode.

The efficiencies of the polarizer and analyzer must be determined prior to extracting the flipper parameters. As detailed in Section 2.1, the \ce{^3He} polarization was measured for both the in-situ and off-situ systems. Based on these values, the flipper-off and flipper-on efficiencies of the spin flipper can be derived as follows:
\begin{align}
    \varepsilon_{\text{off}} &= \frac{AP}{(M_1 - M_2)/(M_1 + M_2)}, \\
    \varepsilon_{\text{on}} &= \frac{AP}{(M_4 - M_3)/(M_4 + M_3)}.
\end{align}
Fig.~\ref{fig:efficiency_wavelength} plots the flipper-on efficiency (red) alongside the flipper-off efficiency (black). At a reference wavelength of \SI{0.34}{\angstrom}, the system yields a flipper-on efficiency of \SI{90}{\percent} against a flipper-off of \SI{70}{\percent}. Notably, while the flipper-on performance remains largely wavelength-independent, the flipper-off efficiency suffers pronounced attenuation as the wavelength increases. The observed decline in flipper-off efficiency is primarily attributed to the magnetic zero-field region at the center of the spin flipper. The simulations were restricted to ideal on-axis trajectories; however, for neutrons deviating from the flipper axis, the magnetic field at the geometric center is non-zero. This deviation results in a reduction of polarization flipper-off efficiency, while leaving the flipper-on efficiency largely unaffected. In the energy region above \SI{2}{eV}, the neutron polarization for both the in-situ and off-situ \ce{^3He} systems drops \SI{17.5}{\percent}. For neutron wavelengths below \SI{0.21}{\angstrom}, the error bars for both flipper-on and flipper-off measurements become substantial, which is likely attributable to insufficient measurement time.

To rigorously quantify the experimental asymmetry, we model the neutron capture yield for a polarized beam interacting with the target. Let $N_0$ denote the total neutron flux exiting the $^3$He polarizer and incident on the target. The beam possesses an instantaneous polarization $\mathcal{P} \in [-1, 1]$.

The total neutron-nucleus cross section is decomposed into a spin-independent component, $\sigma$, and a helicity-dependent interaction strength, $x$. The optical thicknesses for neutrons with spins parallel ($+$) and antiparallel ($-$) to the beam quantization axis are given by $\sigma_{\pm} n = \sigma n \pm x$. Here, $n$ represents the target areal density defined as $n = \rho d$, where $\rho$ is the target density and $d$ is the target thickness.

For a mixed spin state with polarization $\mathcal{P}$, the statistical weights for the parallel and antiparallel states are $w_{\pm}(\mathcal{P}) = (1 \pm \mathcal{P})/2$. The detector count rate $N(\mathcal{P})$, which is proportional to the neutron capture probability, is derived as the weighted sum of the absorption yields:
\begin{align}
    N(\mathcal{P}) &= N_0 \left[ w_+(\mathcal{P}) (1 - e^{-\sigma_+ n}) + w_-(\mathcal{P}) (1 - e^{-\sigma_- n}) \right] \nonumber \\
         &= \frac{N_0}{2} \left[ (1 + \mathcal{P})(1 - e^{-(\sigma n + x)}) + (1 - \mathcal{P})(1 - e^{-(\sigma n - x)}) \right].
\end{align}
Utilizing hyperbolic identities, the rate equation simplifies to separate the spin-independent background from the polarization-dependent signal:
\begin{equation}
    N(\mathcal{P}) = N_0 \left[ 1 - e^{-\sigma n} \cosh(x) + \mathcal{P} e^{-\sigma n} \sinh(x) \right].
    \label{eq:general_rate_function}
\end{equation}
Crucially, the positive sign of the $\mathcal{P}$-dependent term reflects the enhanced absorption associated with the positive helicity state.

In the experiment, the neutron helicity is modulated using a spin flipper. We define $f_n$ as the magnitude of the beam polarization ($0 < f_n \le 1$, corresponding to $f_{\text{in}}$ in the previous context). The effective polarization incident on the target is determined by the spin transport efficiency relations defined earlier: $f_{\text{out}} = f_{\text{in}}(2\varepsilon_{\text{off}} - 1)$ for the un-flipped state and $f_{\text{out}} = f_{\text{in}}(1 - 2\varepsilon_{\text{on}})$ for the flipped state.

The effective polarization states $\mathcal{P}$ incident on the target are thus:
\begin{itemize}
    \item \textbf{Flipper Off ($N_+$):} The system is configured to preserve the initial spin state. Using the efficiency relation for the "off" configuration, the effective polarization is $\mathcal{P}_{\text{off}} = f_n(2\varepsilon_{\text{off}} - 1)$. Substituting into Eq.~(\ref{eq:general_rate_function}):
    \begin{equation}
        N_+ \propto 1 - e^{-\sigma n} \cosh(x) + f_n(2\varepsilon_{\text{off}} - 1) e^{-\sigma n} \sinh(x).
    \end{equation}
    
    \item \textbf{Flipper On ($N_-$):} The system reverses the spin state. Using the efficiency relation for the "on" configuration, the effective polarization is $\mathcal{P}_{\text{on}} = f_n(1 - 2\varepsilon_{\text{on}})$. Note that for high efficiency ($\varepsilon_{\text{on}} \to 1$), this yields $\mathcal{P}_{\text{on}} \approx -f_n$. The rate is:
    \begin{equation}
        N_- \propto 1 - e^{-\sigma n} \cosh(x) + f_n(1 - 2\varepsilon_{\text{on}}) e^{-\sigma n} \sinh(x).
    \end{equation}
\end{itemize}

The experimental asymmetry is defined as $A_{\text{meas}} = (N_+ - N_-)/(N_+ + N_-)$. The difference in count rates isolates the helicity-dependent interaction:
\begin{align}
    N_+ - N_- &= N_0 e^{-\sigma n} \sinh(x) \cdot (\mathcal{P}_{\text{off}} - \mathcal{P}_{\text{on}}) \nonumber \\
              &= 2 N_0 f_n e^{-\sigma n} \sinh(x) (\varepsilon_{\text{off}} + \varepsilon_{\text{on}} - 1).
\end{align}

Assuming small asymmetries, the denominator is dominated by the unpolarized background: $N_+ + N_- \approx 2 N_0 [1 - e^{-\sigma n} \cosh(x)]$.

Consequently, the measured asymmetry relates to the ideal physics asymmetry $A_{\text{ideal}}$ via a linear attenuation factor:
\begin{equation}
    A_{\text{meas}} = \underbrace{\left[ \frac{f_n e^{-\sigma n} \sinh(x)}{1 - e^{-\sigma n} \cosh(x)} \right]}_{A_{\text{ideal}}} \cdot (\varepsilon_{\text{off}} + \varepsilon_{\text{on}} - 1).
    \label{eq:final_asymmetry_correction}
\end{equation}
Equation~(\ref{eq:final_asymmetry_correction}) explicitly demonstrates that imperfect spin transport reduces the observed asymmetry by the factor $(\varepsilon_{\text{off}} + \varepsilon_{\text{on}} - 1)$. This correction is essential for the accurate extraction of the parity-violating matrix element from the raw data.

Applying the correction factor derived in Eq.~(\ref{eq:final_asymmetry_correction}) to the raw asymmetry measured at the \SI{0.747}{eV} resonance accounts for the depolarization effects induced by the spin transport system. With the efficiencies $\varepsilon_{\text{off}}$ and $\varepsilon_{\text{on}}$ quantified via [simulation/measurement], the corrected parity-violating asymmetry is determined to be: 
\begin{equation} 
A_{\text{PV}} = 7.8 \pm 2.4~\text{(stat.)} \pm 0.3~\text{(sys.)}\,\si{\percent}. 
\end{equation}
The statistical uncertainty is propagated from the errors associated with the spin flipper efficiency and the raw asymmetry extraction, while the systematic uncertainty is derived from the analysis of the \SI{70}{eV} resonance.

\subsection{Verification of Polarization Transmission and Its Impact on Measurements}

\begin{figure}[H] 
    \centering
    \includegraphics[width=0.9\linewidth]{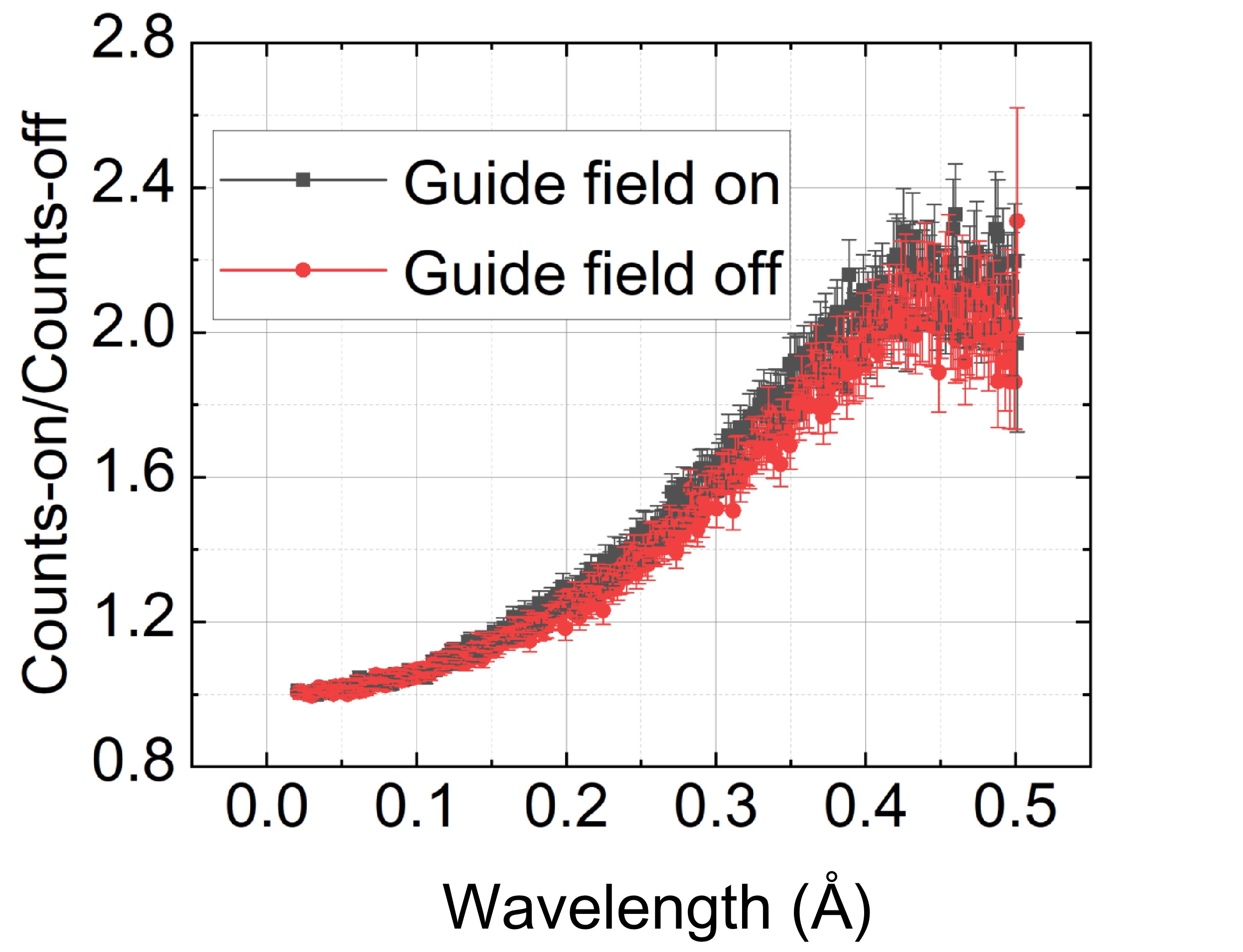} 
    \caption{Wavelength dependence of the measured flipping ratio, defined as the ratio of neutron counts in the flipper-off state to the flipper-on state. The comparison between the datasets with the \SI{1}{m} guide field energized (\textbf{black squares}) and de-energized (\textbf{red circles}) illustrates the stabilizing effect of the guide field on spin transport.}
    \label{fig:flipping_ratio}
\end{figure}

Given the approximately \SI{1.2}{m} distance from the upstream guide field exit to the center of GTAF, a specific test was conducted with the \SI{1}{m} guide coil deactivated to evaluate the impact of this flight path on epithermal neutron polarization transport and the validity of asymmetry measurements. To isolate the contribution of the guide field, comparative neutron data were acquired over two-hour integration periods with the longitudinal solenoid energized and de-energized. Throughout these measurements, instrumental stability was strictly maintained, with the polarizer, spin flipper, analyzer, and \ce{^6Li} detector configurations kept at constant settings. In Fig.~\ref{fig:flipping_ratio} Comparisons between the two datasets reveal that the absence of the \SI{1}{m} guide field leads to only a slight reduction in the flipping ratio for wavelengths above \SI{0.15}{\angstrom}. This indicates that although unshielded ambient magnetic fields induce minor perturbations to the neutron spin trajectory, they are insufficient to fundamentally compromise the beam polarization or the validity of the asymmetry measurements. Nevertheless, to further optimize flipper-off efficiency and minimize potential systematic uncertainties, this active magnetic guidance will be implemented in future experimental campaigns.

\section{Summary and outlook}\label{sec:5}

\begin{figure}[H] 
    \centering
    \includegraphics[width=0.9\linewidth]{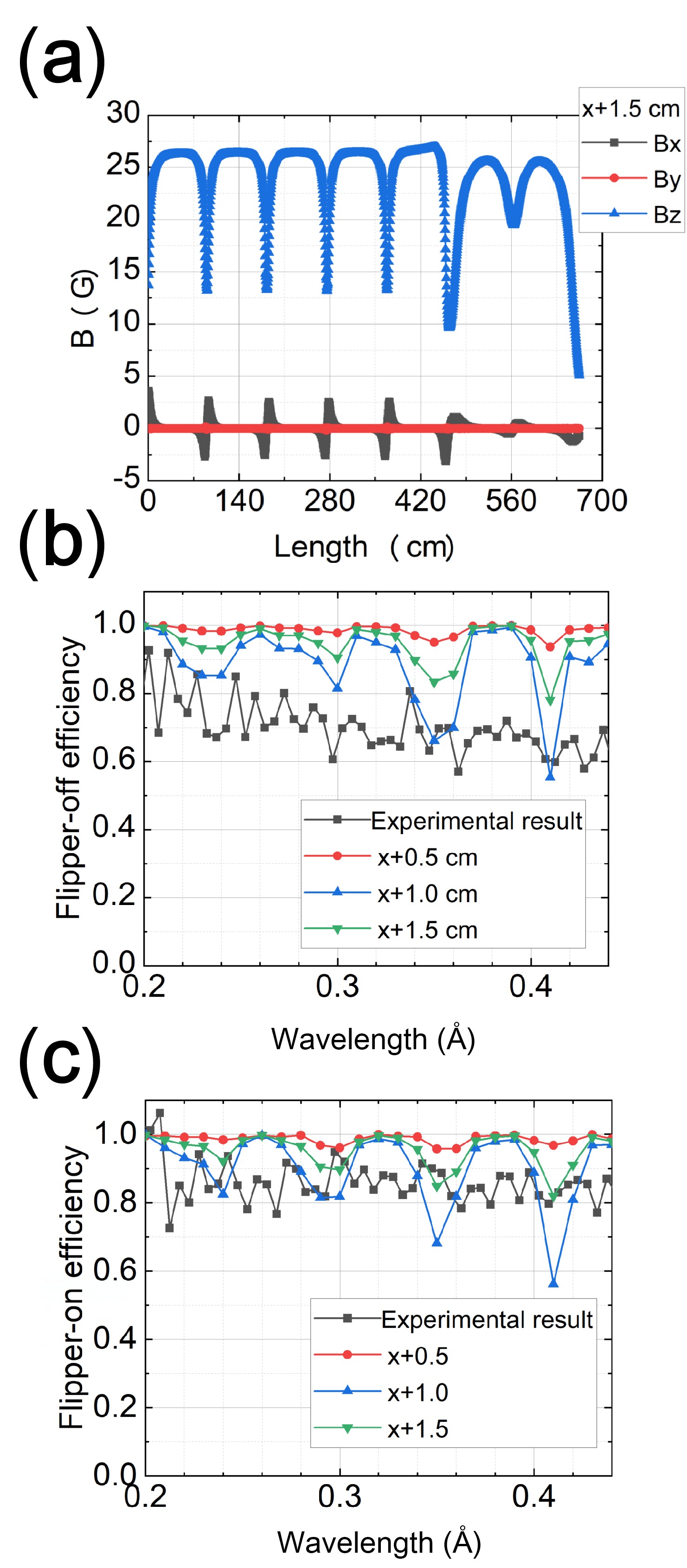} 
    \caption{
    \textbf{(a)} Magnetic field mapping at a distance of 1.5~cm from the beam center with the polarization transport guide field. Comparison of the system performance at the exit of the transport guide for radial offsets of 0.5, 1.0, and 1.5~cm against experimental measurements. 
    \textbf{(b)} Neutron transmission efficiency. 
    \textbf{(c)} Spin flipper-on efficiency.
    }
    \label{fig:GF efficiency}
\end{figure}

As shown in Fig.~\ref{fig:GF efficiency}, the experimentally measured efficiencies are lower than the theoretical values calculated for the standalone spin flipper design. This reduction is primarily attributed to the geometric constraints of the downstream polarization transport system. Spanning a total length of \SI{6.5}{m}, this system requires a segmented guide coil assembly; however, the restricted internal volume of the collimator limits the coil diameter to a maximum of \SI{60}{\mm}. Consequently, the magnetic field lines exhibit curvature away from the central axis, inducing Larmor precession and subsequent depolarization for neutrons with off-axis trajectories. Despite this deviation from the design ideal, the device characterization was performed for the integrated system, coupling the spin flipper with the transport line. As the effective efficiency has been explicitly calibrated as a function of neutron wavelength (see Fig.~\ref{fig:efficiency_wavelength}), these data allow for precise systematic corrections. Therefore, the reduced raw efficiency does not compromise the final accuracy of the upcoming Parity Violation and TRIV measurements.

We have established an eV-energy polarized neutron capability at the CSNS Back-n beamline to support the NOPTREX collaboration's search for Time-Reversal Invariance Violation (TRIV). Leveraging the superior time-of-flight resolution of the unmoderated Back-n source, this system enables Parity Violation (PV) measurements. A longitudinal asymmetry of $7.8\% \pm 2.4\%$ was successfully measured at the \SI{0.74}{eV} $p$-wave resonance of \ce{^{139}La}, validating the system's performance. The apparatus features an in-situ SEOP-based \ce{^3He} neutron polarizer, an adiabatic spin flipper, a solenoid-based magnetic transport system, and a GTAF detector. To minimize systematic errors, the adiabatic spin flipper operates with a custom switching sequence. This configuration yielded a flipper-on efficiency of \SI{90}{\percent} and a flipper-off efficiency of \SI{70}{\percent} at the \SI{0.747}{eV} resonance. The modular design of the system also facilitates future expansions. To further enhance the statistical sensitivity and the practicality of high-precision measurements, we aim to enlarge the beam spot diameter, refine the spin-flip efficiency, and upgrade the performance of the \ce{^3He} spin filter. For example, we plan to extend the capabilities to the thermal energy range by optimizing the pressure and length of the \ce{^3He} cell, and to integrate a vertical cryostat within the GTAF to cool samples below \SI{15}{K}.

% ============================================================================
%   Acknowledgements & References
% ============================================================================

\section*{Acknowledgements}
This work is Supported by National Key R\&D Program of China (No. 2024YFE0110000). The adiabatic flipper development work in this paper was supported by the Guang Dong Basic and Applied Basic Research Foundation Grant No.2021B1515140016. This work was supported by the National Science Fund for Distinguished Young Scholars (Grant No. 12425512), The Back-n WNS is supported by National Key Research and Development Program of China (No.2023YFA1606602). This work was supported by the National Key Research and Development Program of China (Grant No. 2020YFA0406000 and No. 2020YFA0406004). This work is supported by the Guangdong Basic and Applied Basic Research Foundation (Grant No. 2019B1515120079) This work was also supported by National Natural Science Foundation of China (No. 12075265 and No. U2032219) and Guang Dong Basic and Applied Basic Research Foundation (Grant No. 2021B1515140016). The 3He spin filter implemented in the experiment was developed within the Guangdong Provincial Key Laboratory of Extreme Conditions: 2023B1212010002 and the Dongguan Introduction Program of Leading Innovative and Entrepreneurial Talents (No. 20191122). M. Zhang, S. Samiei and W. M. Snow acknowledge support from the US National Science Foundation grant PHY-2209481 and from the Indiana University Center for Spacetime Symmetries. The authors are listed in alphabetical order.

\section*{Conflict of Interest}
The authors declare that they have no conflict of interest.

% Note: arXiv usually prefers standard styles like 'unsrt' or 'apsrev'.
% If you don't upload 'scpma.bst', please change the style below to 'unsrt'.
\bibliographystyle{unsrt} 
\bibliography{references} 

\end{document}